\documentclass[11pt,draftcls, onecolumn, journal,compsoc]{IEEEtran}

\usepackage{amsmath, amssymb,amsthm}
\usepackage{stmaryrd}
\usepackage{epsfig,calc, graphicx}
\usepackage{wasysym}
\usepackage[usenames]{color}

\title{On the Identifiability of Overcomplete Dictionaries via the Minimisation Principle Underlying K-SVD}

\author{Karin Schnass
\thanks{Karin Schnass is with the Computer Vision Laboratory, University of Sassari, Porto Conte Ricerche, 07041 Alghero, Italy, 
	email: kschnass@uniss.it }
}

\newcommand\ip[2]{\langle #1, #2\rangle}
\newcommand\natoms{K}
\newcommand\nsig{N}
\newcommand\sparsity{S}
\newcommand\ddim{d}
\newcommand\ie{{i.e. }}

\newcommand\eps{\varepsilon}
\newcommand\epsmax{\varepsilon_{\max}}
\newcommand\epsmin{\varepsilon_{\min}}
\newcommand\dico{\Phi}
\newcommand\atom{\phi}

\newcommand\pdico{\Psi}
\newcommand\ppdico{\bar\Psi}
\newcommand\patom{\psi}

\newcommand\amp{c}
\newcommand\Noise{R}
\newcommand\noise{r}
\newcommand\nmax{\rho_{\max}}
\newcommand\nsigma{\rho}
\newcommand\An{A_\noise}
\newcommand\net{\mathcal{N}}

\newcommand\diag{\operatorname{diag}}

\newcommand\tr{\operatorname{tr}}

\newcommand{\R}{{\mathbb{R}}}

\newcommand{\E}{{\mathbb{E}}}
\newcommand{\I}{{\mathbb{I}}}
\renewcommand{\P}{{\mathbb{P}}}

\theoremstyle{plain}
\newtheorem{theorem}{Theorem}[section]
\newtheorem{goal}[theorem]{Goal}
\newtheorem{lemma}[theorem]{Lemma}
\newtheorem{corollary}[theorem]{Corollary}

\newtheorem{definition}{Definition}[section]

\theoremstyle{remark}
\newtheorem{example}{Example}[section]
\newtheorem{remark}[example]{Remark}

\renewenvironment{proof}{\noindent
{\bf{Proof:} }}
{\hspace*{\fill}$\Box$\vskip1em}

\begin{document}

\maketitle

\begin{abstract}
\noindent This article gives theoretical insights into the performance of K-SVD, a dictionary learning algorithm that has gained significant popularity in practical applications. The particular question studied here is when a dictionary $\dico\in \R^{\ddim \times \natoms}$ can be recovered as local minimum of the minimisation criterion underlying K-SVD from a set of $N$ training signals $y_n =\dico x_n$. A theoretical analysis of the problem leads to two types of identifiability results assuming the training signals are generated from a tight frame with coefficients drawn from a random symmetric distribution. First, asymptotic results showing, that in expectation the generating dictionary can be recovered exactly as a local minimum of the K-SVD criterion if the coefficient distribution exhibits sufficient decay. Second, based on the asymptotic results it is demonstrated that given a finite number of training samples $N$, such that $N/\log N = O(K^3d)$, 
except with probability $O(N^{-Kd})$ there is a local minimum of the K-SVD criterion within distance $O(KN^{-1/4})$ to the generating dictionary.
\end{abstract}

\begin{keywords}
\noindent dictionary learning, sparse coding, sparse component analysis, K-SVD, finite sample size, sampling complexity, dictionary identification, minimisation criterion, sparse representation
\end{keywords}

\section{Introduction}\label{sec:intro}
As the universe expands so does the information we are collecting about and in it. New and diverse sources such as the internet, astronomic observations, medical diagnostics, etc., confront us with a flood of data in ever increasing dimensions and while we have a lot of technology at our disposal to acquire these data, we are already facing difficulties in storing and even more importantly interpreting them. Thus in the last decades high-dimensional data processing has become a very challenging and interdisciplinary field, requiring the collaboration of researchers capturing the data on one hand and researchers from computer science, information theory, electric engineering and applied mathematics, developing the tools to deal with the data on the other hand. One of the most promising approaches to dealing with high-dimensional data so far has proven to be through the concept of sparsity.\\
A signal is called sparse if it 
has a representation or good approximation in a dictionary, i.e. a representation system like an orthonormal basis or frame, \cite{ch03}, such that the number of dictionary elements, also called atoms,
with non-zero coefficients is small compared to the dimension of the space. Modelling the signals as vectors $y \in \mathbb R^\ddim $ and 
the dictionary accordingly as a matrix collecting $\natoms$ normalised atom-vectors as its columns, i.e. $\dico = (\atom_1,\ldots \atom_\natoms), \atom_i \in \mathbb R^\ddim, \|\atom_i\|_2=1$, we have
$$ y \approx \sum_{i\in I} x(i) \atom_i, $$
for a set $I$ of size $S$, i.e. $|I|=S$, which is small compared to
the ambient dimension, i.e. $S\ll d \leq K$.\\
The above characterisation already shows why sparsity provides such an elegant way of dealing with high-dimensional data. No matter the size
of the original signal, given the right dictionary, its size effectively reduces to a small number of non-zero coefficients. For instance the sparsity of natural images in wavelet bases is the fundamental principle
underlying the compression standard JPEG 2000. \\
Classical sparsity research studies two types of problems. The first line of research investigates how to perform the dimensionality reduction algorithmically, i.e. how to find the sparse approximations of a signal given the sparsity inducing dictionary. By now there exists a substantial amount of theory including a vast choice of algorithms, e.g. \cite{damaav97, chdosa98, cosamp, blda08, dadefogu10}, together with analysis about their worst case or average case performance, \cite{Tropp:greed, tr08, scva07, grrascva08}.
The second line of research investigates how sparsity can be exploited for efficient data processing. So it has been shown that sparse signals are very robust
to noise or corruption and can therefore easily be denoised, \cite{doelte06}, or restored from incomplete information.
This second effect is being exploited in the very active research field of {\it compressed sensing}, see \cite{do06cs,carota06, cs}.\\
However, while sparsity based methods have proven very efficient for high-dimensional data processing, they suffer from one common drawback. They all rely on the existence of a dictionary providing sparse representations for the data at hand.\\
The traditional approach to finding efficient dictionaries is through the careful analysis of the given data class, which for instance has led to the development of wavelets, \cite{da92}, and curvelets, \cite{cadedoyi06}, for natural images.
However when faced with a (possibly exotic) new signal class this analytic approach has the disadvantage of requiring too much time and effort. Therefore, more recently, researchers have started to investigate the possibilities of learning the appropriate dictionary directly from the new data class, i.e. given $\nsig$ signals $y_n\in \R^\ddim$, stored as columns in a matrix $Y=(y_1,\ldots, y_\nsig)$ find a decomposition
$$ Y \approx \dico X$$
into a $\ddim \times \natoms$ dictionary matrix $\dico$ with unit norm columns and a $\natoms \times \nsig$ coefficient matrix with sparse columns. Looking at the matrix decomposition we can immediately see that, on top of being the key to sparse data processing schemes, dictionary learning is actually a  powerful data-analysis tool. Indeed within the blind source separation community dictionary learning is known as sparse component analysis (the dictionary atoms are the sparse components) and this data-analysis point of view has been a parallel driving force for the development of dictionary learning.\\
So far the research focus in dictionary learning has been on algorithmic development rather than theoretic analysis. This means that by now there are several dictionary learning algorithms, which are efficient in practice and therefore popular in applications, see \cite{olsfield96, kreutz03, ahelbr06, mabaposa10, yablda09, krra00, sken10} or \cite{rubrel10} for a more complete survey, but only comparatively little theory. Some theoretical insights come from the blind source separation community, \cite{zipe01,gethci05}, and more recently from a set of generalisation bounds for learned dictionaries, \cite{mapo10, vamabr11, megr12, grjebaklse13}, which predict the quality of a learned dictionary for future data, but unfortunately do not directly imply uniqueness of the 'true' dictionary nor guarantee recoverability by an efficient algorithm, 
However, especially to justify the use of dictionary learning as data analysis tool, we need theoretical identification results quantifying the conditions on the dictionary, the coefficient model generating the sparse signals and the number of training signals under which a scheme will be successful. \\
While it is true that for most schemes we do not yet understand their behaviour, there exists a handful of exceptions to this rule, \cite{ahelbr06b, grsc10, gewawrXX, bagrje13, spwawr12}\footnote{For the sake of completeness we also mention (without discussion) some very recent results, developed while this work has been under review, \cite{argemo13, aganne13, aganjaneta13} .}. For these schemes there are known conditions under which a dictionary can be recovered from a given signal class, but unfortunately they all have certain drawbacks limiting their practical applicability. In \cite{ahelbr06b} the authors themselves state that the algorithm is only of theoretical interest because of its computational complexity and also for the $\ell_1$-minimisation principle, suggested in \cite{zipe01, pl07} and studied in \cite{grsc10, gewawrXX,bagrje13}, finding a local minimum is computational sufficiently challenging to prohibit the learning of very high-dimensional dictionaries. Finally,
the ER-SpUD algorithm, \cite{spwawr12}, has the disadvantage that it can only learn a basis, but not an overcomplete dictionary. \\
In this paper we will start bridging the gap between practically efficient and provably efficient dictionary learning schemes, by providing identification results for the minimisation principle underlying K-SVD (K-Singular Value Decomposition), one of the most widely applied dictionary algorithms.\\
K-SVD was introduced by Aharon, Elad and Bruckstein in \cite{ahelbr06} as a generalisation of the K-means clustering process. The starting point for the algorithm is the following minimisation criterion.
Given some signals $Y=(y_1,\ldots ,y_\nsig)$, $y_n \in \R^d$, find
\begin{align}\label{ksvdcriterion}
\min_{\dico \in \mathcal D, X \in \mathcal X_\sparsity} \| Y - \dico X\|_F^2 
\end{align}
for $\mathcal D: = \{\dico = (\atom_1, \ldots, \atom_K), \atom_i \in \R^d, \|\atom_i\|_2=1\}$ and $\mathcal X_\sparsity:= \{X = (x_1, \ldots, x_N), x_n \in \R^K, \|x_n\|_0\leq\sparsity\}$, where $\|x\|_0$ counts the number of non-zero entries of $x$, and $\|\cdot \|_F$ denotes the Frobenius norm.
In other words we are looking for the dictionary that provides on average the best $\sparsity$-term approximation to the signals in $Y$. \\
K-SVD aims to find the minimum of \eqref{ksvdcriterion} by alternating two procedures, a) fixing the dictionary $\dico$ and finding a new close to optimal coefficient matrix $X^{new}$ column-wise, using a sparse approximation algorithm such as (Orthogonal) Matching Pursuit, \cite{Tropp:greed}, or Basis Pursuit, \cite{chdosa98}, and b) updating the dictionary atom-wise, choosing the updated atom $\atom_i^{new}$ to be the left singular vector to the maximal singular value of the matrix having as its columns the residuals $y_n-\sum_{k\neq i} \atom_k x_n(k)$ of all signals $y_n$ to which the current atom $\atom_i$ contributes, i.e. $X_{ni}=x_n(i)\neq 0$. If in every step for every signal the best sparse approximation is found the K-SVD algorithm is guaranteed to find a local minimiser of \eqref{ksvdcriterion}. However because of the non-optimal sparse approximation procedure it can in general not be guaranteed to converge to a local minimiser of \eqref{ksvdcriterion} unless $S=1$ and a greedy algorithm is used, see also the discussion in Section~\ref{sec:discussion}. We will not go further into algorithmic details, but refer the reader to the original paper \cite{ahelbr06} as well as \cite{ahelbr06b}.
Instead we concentrate on the theoretical aspects of the posed minimisation problem.\\
First it will be convenient to rewrite the objective function using the fact that for any signal $y_n$ the best $\sparsity$-term approximation using $\dico$ is given by the largest projection onto a set of $\sparsity$ atoms $\dico_I=(\atom_{i_1}\ldots \atom_{i_\sparsity} )$, i.e.,
\begin{align*}
\min_{\dico \in \mathcal D, X \in \mathcal X_\sparsity} \| Y - \dico X\|_F^2&= \min_{\dico \in \mathcal D} \sum_n \min_{\|x_n\|_0 \leq S} \| y_n - \dico x_n \|_2^2\\
&= \min_{\dico \in \mathcal D} \sum_n \min_{|I| = S} \| y_n - \dico_I \dico_I^\dagger y_n \|_2^2\\
&= \|Y\|_F - \max_{\dico \in \mathcal D} \sum_n \max_{|I| = S} \| \dico_I \dico_I^\dagger y_n \|_2^2,
\end{align*}
where $\dico_I^\dagger$ denotes the Moore-Penrose pseudo inverse of $\dico_I$. Abbreviating the  projection onto the span of $(\atom_i)_{i\in I}$ by $P_I(\dico)=\dico_I \dico_I^\dagger$, we can thus replace the minimisation problem in \eqref{ksvdcriterion} with the following maximisation problem,
\begin{align}\label{maxcriterion}
\max_{\dico \in \mathcal D} \sum_n \max_{|I| = S} \| P_I(\dico) y_n \|_2^2.
\end{align}
From the above formulation it is quite easy to see the motivation for the proposed learning criterion. Indeed assume that the training signals are all $\bar \sparsity$-sparse in an admissible dictionary $\bar\dico \in \mathcal D$, i.e. $Y=\bar{\dico}\bar{X}$ 
and $\|\bar{x}_i\|_0\leq\bar \sparsity$, then
clearly there is a global maximum\footnote{$\bar\dico$ is a global maximiser together with all $2^KK!$ dictionaries consisting of a permutation of the atoms in $\bar\dico$ provided with a $\pm 1$ sign. For a more detailed discussion on the uniqueness of the maximiser/minimiser see eg. \cite{grsc10}.} 
 of~\eqref{maxcriterion} at $\bar{\dico}$, respectively a global minimum of \eqref{ksvdcriterion} at $(\bar{\dico},\bar{X})$, as long as $\bar{\sparsity}\leq\sparsity$. However in practice we will be facing something like,
\begin{align}\label{sigmodabstract}
y_n=\bar{\dico}\bar{x}_n + \noise_n \qquad \mbox {or} \qquad Y= \bar{\dico}\bar{X} + \Noise,
\end{align}
where the coefficient vectors $\bar{x}_n$ in $\bar{X}$ are only approximately $S$-sparse or rapidly decaying
and the pure signals $\bar{\dico}\bar{x}_n $ are corrupted with noise $\Noise =(\noise_1,\ldots,\noise_\natoms)$.
In this case it is no longer trivial or obvious that $\bar{\dico}$ is a local maximum of \eqref{maxcriterion}, but we can hope for a result of the following type.
\begin{goal}\label{th:goal}
Assume that the signals $y_n$ are generated as in \eqref{sigmodabstract}, with $x_n$ drawn from a distribution of approximately sparse or decaying vectors and $r_n$ random noise. As soon as the number of signals $\nsig$ is large enough $\nsig \geq C$, with high probability $p \approx 1$ there will be a local maximum of \eqref{maxcriterion} within distance $\eps$ from $\bar{\dico}$.
\end{goal}
The rest of this paper is organised as follows. After introducing some notation in Section~\ref{sec:notation}, we first give conditions on the dictionary and the coefficients which allow for asymptotic identifiability by studying 
when $\bar{\dico}$ is exactly at a local maximum in the limiting case, where we replace the sum in~\eqref{maxcriterion} with the expectation,
\begin{align}\label{maxexpect}
\max_{\dico \in \mathcal D}\, \E_y\left(\max_{|I| = S} \| P_I(\dico) y\|_2^2\right).
\end{align}
Thus in Section~\ref{sec:asymptotic} we will prove identification results for \eqref{maxexpect} assuming first a simple (discrete, noise-free) signal model and then progressing to a noisy, continuous signal model. In Section~\ref{sec:finite} we will go from asymptotic results to results for finite sample sizes and prove versions of Theorem~\ref{th:goal} that under the same assumptions as the asymptotic results quantify the sizes of the parameters $\eps, p$ in terms of the number of training signals $\nsig$ and the size of $C$ in terms of the number of atoms $\natoms$. In the last section we will discuss the implications of our results for practical applications, compare them to existing identification results and point out some directions for future research.

\section{Notations and Conventions}\label{sec:notation}
Before we jump into the fray, we collect some definitions and lose a few words on notations; usually subscripted letters will denote vectors with the exception of $\amp$ and $\eps$ where they are numbers, eg. $ (x_1,\ldots,x_\natoms)=X \in \R^{d\times \natoms}$ vs. $\amp=(\amp_1, \ldots, \amp_\natoms)\in \R^\natoms$, however, it should always be clear from the context what we are dealing with. \\
For a matrix $M$, we denote its (conjugate) transpose by $M^\star$ and its Moore-Penrose pseudo inverse by $M^\dagger$. We denote its operator norm by $\|M\|_{2,2}=\max_{\|x\|_2=1}\|Mx\|_2$ and its Frobenius norm by $\|M\|_F= \tr(M^\star M)^{1/2}$, remember that we have $\|M\|_{2,2}\leq \|M\|_F$.\\
We consider a {\bf dictionary} $\dico$ a collection of $K$ unit norm vectors $\atom_i\in \R^d$, $\|\atom_i\|_2=1$. By abuse of notation we will also refer to the $d \times K$ matrix collecting the atoms as its columns as the dictionary, i.e. $\dico=(\atom_i, \ldots \atom_K)$. The maximal absolute inner product between two different atoms is called the {\bf coherence} $\mu$ of a dictionary, $\mu=\max_{i \neq j}|\ip{\atom_i}{\atom_j}|$.\\
By $\dico_I$ we denote the restriction of the dictionary to the atoms indexed by $I$, i.e. $\dico_I=(\atom_{i_1}\ldots \atom_{i_\sparsity} )$, $i_j\in I$, and by $P_I(\dico)$ the orthogonal projection onto the span of the atoms indexed by $I$, i.e. $P_I(\dico)=\dico_I \dico_I^\dagger$. Note that in case the atoms indexed by $I$ are linearly independent we have $\dico_I^\dagger = (\dico_I^\star \dico_I)^{-1} \dico_I^\star$.\\
(Ab)using the language of compressed sensing we denote the minimal eigenvalue of $\dico^\star_I\dico_I$ by $1-\delta_I(\dico)$ and define the {\bf lower isometry constant} $\delta_S(\dico)$ of the dictionary as $\delta_S(\dico):=\max_{|I|\leq S} \delta_I(\dico)\leq 1$. If any set of $S$ atoms is linearly independent we have $\delta_S(\dico)<1$ and in general we have the bound $\delta_S(\dico) \leq \mu (S-1)$. When clear from the context we will usually omit the reference to the dictionary. For more details on isometry constants, see for instance \cite{carota06}.\\
For two dictionaries $\dico,\pdico$ we define the distance between each other as the maximal distance between two corresponding atoms, i.e.
\begin{align}
d(\dico,\pdico):=\max_i \|\atom_i-\patom_i\|_2.
\end{align}
We consider a {\bf frame} $F$ a collection of $K\geq d$ vectors $f_i\in\R^\ddim$ for which there exist two positive constants $A,B$ such that for all $v \in \R^\ddim$ we have
\begin{align}\label{framebound}
A \|v\|^2_2 \leq \sum_{i=1}^K |\ip{f_i}{v}|^2 \leq B \|v\|^2_2.
\end{align}
If $B$ can be chosen equal to $A$, i.e. $B=A$, the frame is called tight and if all elements of a tight frame have unit norm we have $A=K/\ddim$. The operator $FF^\star$ is called frame operator and by \eqref{framebound} its spectrum is bounded by $A, B$. For more details on frames, see e.g. \cite{ch03}.\\
Finally we introduce the Landau symbols $O,o$ to characterise the growth of a function. We write $f(\eps)=O(g(\eps))$ if $\lim_{\eps \rightarrow 0} f(\eps)/g(\eps)= C<\infty$ and $f(\eps)=o(g(\eps))$ if $\lim_{\eps \rightarrow 0} f(\eps)/g(\eps)=0$.

\section{Asymptotic identification results \label{sec:asymptotic}}

As mentioned in the introduction if the signals $y$ are all $S$-sparse in a dictionary $\bar\dico$ then clearly there is a global minimum of~\eqref{ksvdcriterion} or global maximum of~\eqref{maxexpect} with parameter $S$ at $\bar \dico$. However what happens if we do not have perfect $S$-sparsity? Let us start with a very simple negative example of a coefficient distribution for which the original generating dictionary is not at a local maximum for the case $S=1$.

\begin{example}\label{ex:flatcoeff}
Let $U$ be an orthonormal basis and let the signals be generated as $y=U x$, where $x$ is a randomly 2-sparse, 'flat' coefficients sequence, i.e. we pick an index set $I=\{i,j\}$ and two signs $\sigma_{i/j}=\pm 1$ uniformly at random and set $x(k)=\sigma_k$ for $k\in I $ and zero else. Then there is no local maximum of~\eqref{maxexpect} with $S=1$ at $U$. Indeed since the signals are all 2-sparse the maximal inner product with all atoms in $U$ is the same as the maximal inner product with only $d-1$ atoms. This degree of freedom we can use to construct an ascent direction. Choose $U_\eps=(u_1, \ldots, u_{d-1}, (u_d + \eps u_1)/\sqrt{1+\eps^2})$. Using the identity $\max_{i} \| P_i(\dico) y\|_2^2 = \| \dico^\star y\|_\infty^2$ we get,
\begin{align*}
\E_y\left( \| U_\eps^\star y\|_\infty^2\right)&=\E_x \left( \| U_\eps^\star U x \|_\infty^2\right)\\
&= \E_x\left(\| (x(1), \ldots, x(d-1), {\textstyle \frac{x(d) + \eps x(1)}{\sqrt{1+\eps^2}}}) \|_\infty^2\right)\\
&=1 \cdot (1 -  \P (I = \{1, d\}\cap \sigma_1= \sigma_d) ) + \frac{(1+\eps)^2}{1+\eps^2} \cdot \P (I = \{1, d\} \cap \sigma_1=\sigma_d) \\
&= 1 + \frac{\eps}{1+\eps^2}\cdot \frac{1}{d(d-1)},
\end{align*}
which is larger than $\E_y\left(\| U^\star y\|_\infty^2\right)=1$.
\end{example}
From the above example we see that in order to have a local maximum at the original dictionary we need a signal/coefficient model where the coefficients show some type of decay. 

\subsection{A simple model of decaying coefficients}\label{sec:simplemodel}
To get started we consider a very simple coefficient model, constructed from a non-negative, non-increasing sequence $\amp \in \R^\natoms$ with $\|c\|_2=1$, which we permute uniformly at random and provide with random $\pm$ signs. To be precise for a permutation $p:\{1,...,\natoms\}\rightarrow\{1,...,\natoms\}$ 
and a sign sequence $\sigma$, $\sigma_i=\pm 1$, we define the sequence $\amp_{p, \sigma}$ component-wise as
$\amp_{p, \sigma}(i):=\sigma_i \amp_{p(i)}$, and set $y=\dico x$ where $x=\amp_{p, \sigma}$ with probability $(2^\natoms \natoms!)^{-1}$. \\
The normalisation $\|c\|_2=1$ has the advantage that for dictionaries, which are an orthonormal basis,
the resulting signals also have unit norm and for general dictionaries the signals have unit square norm in expectation, i.e. $\E (\|y\|_2^2)=1$. This reflects the situation in practical applications, where we would normalise the signals in order to equally weight their importance.\\
Armed with this model we can now prove a first dictionary identification result for~\eqref{maxexpect}.
%
%
\begin{theorem}\label{th:simplemodel}
Let $\dico$ be a unit norm tight frame with frame constant $A=\natoms/\ddim$ and lower isometry constant $\delta_S$.
Let $x$ be a random permutation of a positive, nonincreasing sequence $\amp$, where $\amp_1 \geq \amp_2 \geq \amp_3 \ldots \geq \amp_\natoms \geq 0$ and $\|\amp\|_2=1$, provided with random $\pm$ signs, \ie $x=\amp_{p, \sigma}$ with probability $\P(p,\sigma)=(2^\natoms \natoms!)^{-1}$. Assume that the signals are generated as $y=\dico x$.
If there exists $\kappa>0$ such that for $I_p:=p^{-1}\left(\{1,\ldots \sparsity\}\right)$ we have
\begin{align}\label{maxatIp}
\|P_{I_p}(\dico)\dico \amp_{p,\sigma}\|_2 - \max_{|I| = S,I\neq I_p}\|P_{I}(\dico)\dico \amp_{p,\sigma}\|_2 \geq 2\kappa, \quad \forall \sigma, p,
\end{align}
then there is a local maximum of \eqref{maxexpect} at $\dico$. \\
Moreover for $\pdico\neq \dico$ we have $\E_y\left(\max_{|I| = S} \| P_I(\pdico) y\|_2^2\right) < \E_y\left(\max_{|I| = S} \| P_I(\dico) y\|_2^2\right)$ as soon
as 
\begin{align}\label{eq:epsmax}
d(\dico,\pdico) \leq  \frac{\kappa \sqrt{1-\delta_S} }{\sqrt{\frac{9S}{2}}\left(1+ 4\sqrt{\log\left(\frac{60 A {K \choose S}^2}{ \kappa \lambda_S (1-\delta_S) }\right)} \right)},
\end{align}
where $\lambda_S =  \frac{\amp_1^2+\ldots+\amp_\sparsity^2}{S}-\frac{1-\amp_1^2-\ldots-\amp_\sparsity^2}{K-S} $ and $\delta_S <1$ because of \eqref{maxatIp}.
\end{theorem}

\begin{proof}
The basic idea of the proof is that for the original dictionary the maximal response is always attained for the set $I_p$ and that for most signals, i.e. most sign sequences, also for a perturbed dictionary the maximal response is still at $I_p$. Since the average loss of a perturbed dictionary over most sign sequences,
\begin{align}
\E_p \E_\sigma \left(  \| P_{I_p}(\pdico) \dico\amp_{p,\sigma}\|_2^2 \right) < \E_p \E_\sigma \left(  \| P_{I_p}(\dico) \dico\amp_{p,\sigma}\|_2^2 \right),
\end{align} is larger than the possible gain on exceptional sign sequences we have a maximum at $\dico$. More detailed sketches and a version of the proof for $S=1$ can be found in \cite{sc13arxiv, sc13sampta}.\\
Following the proof idea we first calculate the expectation using the original dictionary $\dico$. Condition~\eqref{maxatIp} quite obviously (and artlessly) guarantees that the maximum is always attained for the set $I_p$, so setting $\gamma_S^2:=\amp_1^2+\ldots+\amp_\sparsity^2$ we get from Lemma~\ref{lem:expect} in the appendix,
\begin{align}
\E_y\left(\max_{|I| = S} \| P_I(\dico) y\|_2^2\right)&=\E_p \E_\sigma \left(  \| P_{I_p}(\dico) \dico\amp_{p,\sigma}\|_2^2 \right)\notag \\
&= \frac{A(1-\gamma_S^2)S}{(K-S)} + \left(\frac{\gamma_S^2}{S}-\frac{1-\gamma_S^2}{K-S}\right){K \choose S}^{-1}\sum_{I: |I|=S} \| \dico_I\|_F^2 \label{eq:expatdico}.
\end{align}
To compute the expectation for a perturbation of the original dictionary we
first note that we can parametrise all $\eps$-perturbations $\pdico$ of the original dictionary $\dico$, i.e. $d(\dico,\pdico)=\eps$, as
$$
\patom_i = (1-\eps^2_i/2) \atom_i + (\eps_i^2 - \eps_i^4/4)^{\frac{1}{2}}  z_i,
$$
for some $z_i$ with $\langle \atom_i,z_i\rangle = 0, \|z_i\|_2=1$ and some $\eps_i$ with 
$\max_i \eps_i = \eps$. For conciseness of the following presentation we define 
$\alpha_i := 1-\eps^2_i/2$, $\omega_i := (\eps_i^2 - \eps_i^4/4)^{\frac{1}{2}}$ and $b_i: = \omega_i/\alpha_i z_i$. 
Further we define $A_I=\diag(\alpha_i)_{i\in I}$ and $W_I=\diag(\omega_i)_{i\in I}$ to get $\pdico_I=\dico_I A_I + Z_I W_I$ and $B_I= Z_I W_IA_I^{-1}$. Note that some perturbations, e.g. small rotations, will be also unit norm tight frames but in general the perturbed dictionaries will not be tight.\\
As pointed out in the proof idea our strategy will be to show that for a fixed permutation $p$ with high probability (over $\sigma$) the maximal projection is still onto the atoms indexed by $I_p$.\\
For any index set $I$ of size $\sparsity$ we can bound the projection onto a perturbed dictionary as, 
\begin{align} \label{eq:worstcase}
 \| P_I(\pdico) y\|^2_2&  =  \| P_I(\dico) y\|^2_2 + \ip{P_I(\dico) y}{\big(P_I(\pdico) - P_I(\dico)\big) y} + \ip{P_I(\pdico) y}{\big(P_I(\pdico) - P_I(\dico)\big) y}  \notag\\
&\leq \| P_I(\dico) y\|^2_2 + 2 \| y \|_2 \|\big(P_I(\pdico) - P_I(\dico)\big) y\|_2 \notag \\
&\leq \| P_I(\dico) y\|^2_2 + 2 \| y \|^2_2 \| \| P_I(\pdico) - P_I(\dico)\|_{2,2} \notag \\
& \leq  \| P_I(\dico) y\|^2_2 +2A \max_{|I| = S}\|P_I(\pdico) - P_I(\dico)\|_{F},
\end{align}
leading to 
\begin{align} \label{pessimistic}
\max_{|I| = S} \| P_I(\pdico) y\|^2_2 \leq  \| P_{I_p}(\dico) y\|^2_2 +2A \max_{|I| = S}\|P_I(\pdico) - P_I(\dico)\|_{F}.
\end{align}
However \eqref{pessimistic} is a quite pessimistic estimate since for most $y= \dico \amp_{p,\sigma}$, meaning for most $\sigma$, the expression $\|\big(P_I(\pdico) - P_I(\dico)\big) y\|_2$ will be much smaller than $\| P_I(\pdico) - P_I(\dico)\|_{2,2} \| y\|_2$. Indeed we can estimate its typical size via the following convenient if not optimal concentration inequality for Rademacher series from \cite{leta91}, Chapter 4. 
\begin{corollary}[of Theorem~4.7 in \cite{leta91}]
For a vector-valued Rademacher series $V=\sum_i \sigma_i v_i$, i.e. for $\sigma_i$ independent Bernoulli variables with $\P(\sigma_i=\pm1)=1/2$ and $v_i \in \R^n$, and $t>0$ we have,
\begin{align}
\P(\|V\|_2>t)\leq 2\exp\left(\frac{-t^2}{32\E(\|V\|_2^2)} \right).
\end{align}
\end{corollary}
Applied to $v_i = \amp_{p(i)}\big(P_I(\pdico) - P_I(\dico)\big)\atom_i$ this leads to the following estimate,
\begin{align}
\P\left( \|\big(P_I(\pdico) - P_I(\dico)\big)\dico \amp_{p,\sigma}\|_2>t \right)&
\leq 2\exp \left(\frac{-t^2}{32\sum_i \amp_{p(i)}^2 \| \big(P_I(\pdico) - P_I(\dico)\big)\atom_i\|^2_2} \right)\notag \\
&
\leq 2\exp \left(\frac{-t^2}{32\sum_i \amp_{p(i)}^2 \| P_I(\pdico) - P_I(\dico)\|^2_{2,2}} \right)\notag \\
&
\leq 2\exp \left(\frac{-t^2}{32 \| P_I(\pdico) - P_I(\dico)\|^2_F} \right), \label{projbound}
\end{align}
whenever $P_I(\pdico) \neq P_I(\dico)$ - otherwise we trivially have $\P\left( \|\big(P_I(\pdico) - P_I(\dico)\big)\dico \amp_{p,\sigma}\|_2>t \right)=0$. 
We now define the set $\Sigma_p$,
\begin{align}\label{defSigmap}
\Sigma_p := \bigcup_{I: |I|=\sparsity} \{\sigma:  \| \big(P_I(\pdico) - P_I(\dico)\big)\dico \amp_{p,\sigma}\|_2>\kappa\},
\end{align}
whose size we can estimate using \eqref{projbound} with $t=\kappa$ and a union bound, 
\begin{align}
\P(\Sigma_p)\leq 2 \sum_{I: P_I(\pdico) \neq P_I(\dico)} \exp \left(\frac{-\kappa^2}{32 \| P_I(\pdico) - P_I(\dico)\|^2_F}\right):=\eta_S.
\end{align}
Note that whenever $\sigma \notin \Sigma_p$ we have $\max_{I } \|P_{I}(\pdico)\dico \amp_{p,\sigma}\|_2 = \|P_{I_p}(\pdico)\dico \amp_{p,\sigma}\|_2$, since using the (reversed) triangular inequality we have
\begin{align}
\|P_{I_p}(\pdico)\dico \amp_{p,\sigma}\|_2& \geq \|P_{I_p}(\dico)\dico \amp_{p,\sigma}\|_2 -  \|\big(P_I(\pdico) - P_I(\dico)\big)\dico \amp_{p,\sigma}\|_2 \notag \\
& \geq \|P_{I_p}(\dico)\dico \amp_{p,\sigma}\|_2 -  \kappa \notag\\
&\geq \max_{I : I\neq I_p} \|P_{I}(\dico)\dico \amp_{p,\sigma}\|_2 + \kappa \notag\\
&\geq \max_{I : I\neq I_p} \left(\|P_{I}(\dico)\dico \amp_{p,\sigma}\|_2+\|\big(P_I(\pdico) - P_I(\dico)\big)\dico \amp_{p,\sigma}\|_2 \right)\notag \\
& \geq 
\max_{I : I\neq I_p} \|P_{I}(\pdico)\dico \amp_{p,\sigma}\|_2.
\end{align}
To finally calculate the expectation over $\sigma$ for a perturbed dictionary we split it into a sum over the sign sequences contained in $\Sigma_p$ and its complement. We can estimate, 
\begin{align}
 \E_\sigma& \left(\max_{|I| = S} \| P_I(\pdico) \amp_{p,\sigma}\|_2^2\right)\notag \\
&= \sum_{\sigma \in \Sigma_p} \max_{|I| = S} \| P_I(\pdico) \dico \amp_{p,\sigma}\|_2^2 +  \sum_{\sigma \notin \Sigma_p}\max_{|I| = S} \| P_I(\pdico) \dico \amp_{p,\sigma}\|_2^2\notag \\
&\leq \sum_{\sigma \in \Sigma_p}\left(\| P_{I_p}(\dico) \amp_{p,\sigma}\|^2_2 +2A \max_{|I| = S}\|P_I(\pdico) - P_I(\dico)\|_{F} \right)+  \sum_{\sigma \notin \Sigma_p}\max_{|I| = S} \| P_I(\pdico) \dico \amp_{p,\sigma}\|_2^2 \label{eq:splitsum1}\\
&\leq \sum_{\sigma \in \Sigma_p}\left(\| P_{I_p}(\pdico) \amp_{p,\sigma}\|^2_2 +4A \max_{|I|= S}\|P_I(\pdico) - P_I(\dico)\|_{F} \right)+  \sum_{\sigma \notin \Sigma_p} \| P_{I_p}(\pdico) \dico \amp_{p,\sigma}\|_2^2 \label{eq:splitsum2}\\
 &\leq 4 \eta_\sparsity A\max_{|I| = S} \|P_I(\pdico) - P_I(\dico)\|_{F} + \E_\sigma\left( \| P_{I_p}(\pdico) \dico \amp_{p,\sigma}\|_2^2\right), \notag
\end{align}
where we have used \eqref{eq:worstcase}, reversing the roles of $\dico$ and $\pdico$ and choosing $I=I_p$, to go from \eqref{eq:splitsum1} to \eqref{eq:splitsum2}.
Using the expression for $\E_p\E_\sigma\left( \| P_{I_p}(\pdico) \dico \amp_{p,\sigma}\|_2^2\right)$ derived in Lemma~\ref{lem:expect} in the appendix we get the following bound for the expectation of the maximal projection using a perturbed dictionary,
\begin{align}
\E_p \E_\sigma \left(\max_{|I| = S} \| P_I(\pdico)\dico \amp_{p,\sigma}\|_2^2\right)
 &\leq   
 \frac{A(1-\gamma_S^2)S}{(K-S)} + \left( \frac{\gamma_S^2}{S}-\frac{1-\gamma_S^2}{K-S}\right) {K \choose S}^{-1}\sum_{I:|I|=S} \| P_I(\pdico) \dico_I\|_F^2\notag \\
 &\hspace{4cm} +  4 \eta_\sparsity A\max_{|I|=S}  \|P_I(\pdico) - P_I(\dico)\|_{F}\label{eq:expatpdico} .
 \end{align}
We are now ready to compare the above expression to the corresponding one for the original dictionary. Abbreviating $\lambda_S =   \frac{\gamma_S^2}{S}-\frac{1-\gamma_S^2}{K-S} $ 
and using the estimates for $\|P_I(\pdico) - P_I(\dico)\|_{F}$ and $\| P_I(\pdico) \dico_I\|_F^2-\| \dico_I\|_F^2$ from Lemma~\ref{lem:projpdico} in the appendix, we get 
\begin{align}
\E_y &\left(\max_{|I| =S} \| P_I(\pdico)y\|_2^2\right) - \E_y \left(\max_{|I| = S} \| P_I(\dico)y\|_2^2\right) \notag \\
 &\qquad \leq 4 A\max_{|I|=S}  \|P_I(\pdico) - P_I(\dico)\|_{F} \sum_{P_I(\pdico) \neq P_I(\dico)} \exp \left(\frac{-\kappa^2}{32 \| P_I(\pdico) - P_I(\dico)\|^2_F} \right) \notag \\
 &\hspace{8cm} +  \lambda_S { {K \choose S}^{-1}} \sum_{I:|I|=S} \left( \| P_I(\pdico) \dico_I\|_F^2-\| \dico_I\|_F^2 \right)\notag\\
 &\qquad\leq \frac{4 A C_1}{\sqrt{1-\delta_S}}\max_{|I|=S}  \| Q_I(\dico)B_I  \|_F \sum_{I:Q_I(\dico)B_I\neq 0} \exp \left(\frac{-\kappa^2 (1-\delta_S)}{32 C^2_1 \| Q_I(\dico)B_I  \|^2_F} \right) \notag \\
   &\hspace{8cm} -   \lambda_S { {K \choose S}^{-1}} \sum_{I:|I|=S} C_2 \| Q_I(\dico)B_I  \|^2_F,
   \end{align}
 with $C_1=1.487$ and $C_2=0.897$ and where we have used that \eqref{eq:epsmax} implies $\eps \leq \frac{\sqrt{1-\delta_S}}{21\sqrt{S}}$. Denote by $\bar I $ the set for which $ \| Q_I(\dico)B_I  \|_F$ is maximal. 
 We can further estimate,
\begin{align}
\E_y &\left(\max_{|I| = S} \| P_I(\pdico)y\|_2^2\right) - \E_y \left(\max_{|I| = S} \| P_I(\dico)y\|_2^2\right) \notag \\
  &\qquad\leq \frac{4 A C_1}{\sqrt{1-\delta_S}}  \| Q_{\bar I}(\dico)B_{\bar I}  \|_F   {K \choose S} \exp \left(\frac{-\kappa^2 (1-\delta_S)}{32 C^2_1  \| Q_{\bar I}(\dico)B_{\bar I}  \|_F^2} \right) -   \lambda_S { {K \choose S}^{-1}} C_2 \| Q_{\bar I}(\dico)B_{\bar I}  \|_F^2. \notag
\end{align}
Thus to have a local maximum at $\dico$ we need to show that for $\eps \neq 0$ small enough we have
\begin{align}
 \frac{4 A C_1}{\sqrt{1-\delta_S}}  \| Q_{\bar I}(\dico)B_{\bar I}  \|_F   {K \choose S} \exp \left(\frac{-\kappa^2 (1-\delta_S)}{32 C^2_1  \| Q_{\bar I}(\dico)B_{\bar I}  \|_F^2} \right) <  \lambda_S { {K \choose S}^{-1}} C_2 \| Q_{\bar I}(\dico)B_{\bar I}  \|_F^2,\notag
\end{align}
or equivalently that
\begin{align}
 \frac{4 A C_1}{\lambda_S C_2 \sqrt{1-\delta_S}}   {K \choose S}^2 \exp \left(\frac{-\kappa^2 (1-\delta_S)}{32 C^2_1  \| Q_{\bar I}(\dico)B_{\bar I}  \|_F^2} \right) <  \| Q_{\bar I}(\dico)B_{\bar I}  \|_F. 
\end{align}
Applying Lemma~\ref{lem:trick} we get that for $\| Q_{\bar I}(\dico)B_{\bar I}  \|_F > 0 $ the inequality above is satisfied if we have
\begin{align}
\| Q_{\bar I}(\dico)B_{\bar I}  \|_F \leq \frac{4\kappa \sqrt{1-\delta_S} }{C_1\sqrt{32}\left(1+ \sqrt{1+16\log\left(\frac{4\sqrt{32}C^2_1 A {K \choose S}^2}{C_2 \kappa \lambda_S (1-\delta_S) }\right)} \right)}.
\end{align}
Employing the bound $\| Q_{\bar I}(\dico)B_{\bar I}  \|^2_F \leq\| B_{\bar{I}} \|^2_F\leq \sparsity \eps^2/(1-\eps^2)$ this is further implied by 
\begin{align}
\frac{\eps}{\sqrt{1-\eps^2}}\leq \frac{\kappa \sqrt{1-\delta_S} }{C_1\sqrt{2S}\left(1+ 4\sqrt{\log\left(\frac{4\sqrt{32}e^{1/16} C^2_1 A {K \choose S}^2}{C_2 \kappa \lambda_S(1-\delta_S) }\right)} \right)},
\end{align}
which is in turn implied by \eqref{eq:epsmax}.\\
Finally all that remains to show is that for $\eps > 0$ we have $\| Q_{\bar I}(\dico)B_{\bar I}  \|_F > 0 $. Assume conversely that for $\eps>0$ we have $\| Q_{\bar I}(\dico)B_{\bar I}  \|_F = 0 $ meaning that $\| Q_{ I}(\dico)B_{ I}  \|_F = 0 $ for all $I$ of size $S$. We can then find an index $\iota $ for which we have
$\patom_\iota = (1-\eps^2/2) \atom_i + (\eps^2 - \eps^4/4)^{\frac{1}{2}}  z_\iota$ for some $z_i$ with $\ip{ \atom_\iota}{z_\iota}=0$ and $\|z_\iota\|_2$=1. For all $I$ of size $S$ containing $\iota$ we have $Q_I (\dico) b_\iota=0$ and therefore $Q_I (\dico)z_\iota=0$. Choose $J$ to be any set of size $S-1$ containing $\iota$. For all $j\notin J$ we have $Q_{J\cup{j}}(\dico) z_\iota = 0$ or $P_{J\cup{j}}(\dico) z_\iota = z_\iota$, which means that either $z_\iota$ is in the span of $\dico_J$ and therefore $Q_{J}(\dico) z_\iota = 0$ or that
$\atom_j$ is in the span of $(\dico_J, z_i)$ for all $j\notin J$. However this would mean that $\dico$ has rank $S<d$ which is a contradiction to $\dico$ being a frame and we can conclude that $Q_I (\dico)z_\iota=0$ for all $I$ of size $S-1$ containing $\iota$. Iterating the argument we get that $z_\iota$ has to be in the span of $\atom_\iota$ which is a contradiction to $\ip{ \atom_\iota}{z_\iota}=0$ and $\|z_\iota\|_2$=1. 
\end{proof}

\begin{remark}
\noindent (a) To make the theorem more applicable it would be nice to have a concrete condition in terms of the coherence of the dictionary rather than the abstract condition in~\eqref{maxatIp}. Indeed it can be shown, see \cite{sc13arxiv} Appendix C, that we can find a $\kappa>0$ if we have $S\mu <1/2$ and
\begin{align}\label{decaycond}
c_S >  \frac{1-S\mu}{1-2S\mu} c_{S+1} +  \frac{4 \mu}{1-2S\mu} \sum_{i>\sparsity+1} |\amp_i|.
\end{align}
In some cases we can also easily derive estimates for $\kappa$.\\
If $\dico$ is an orthonormal basis we have
\begin{align}\label{kappaonb}
\kappa \geq \frac{c_S^2 - c_{S+1}^2 } {2\sqrt{c_1^2 +\ldots + c_S^2}},
\end{align}
and if $S=1$ we have
\begin{align}\label{kappaonb}
\kappa \geq (c_1-c_2)(1-\mu) - 2\mu \sum^K_{i=3}c_i.
\end{align}
\noindent (b)  Next note that in some sense the theorem is sharp.
Assume that $\dico$ is an orthonormal basis. Then we simply have $\|P_{I}(\dico)\dico \amp_{p,\sigma}\|^2_2=\sum_{i\in I}c_{p(i)}^2$ and the condition to be a local minimum reduces to $c_S> c_{S+1}$. However similar to Example~\ref{ex:flatcoeff} if $c_S= c_{S+1}$ we can again construct an ascent direction and so $\dico$ is not a local maximum.\\
\noindent (c) Finally before extending Theorem~\ref{th:simplemodel} to more general coefficient models we want to motivate why we used the condition that $\dico$ is a tight frame. \\
Assume the same conditions as in Theorem~\ref{th:simplemodel} but that $\dico$ is not tight, i.e. $A\|v\|_2^2\leq \sum_i |\ip{v}{\atom_i}|^2\leq B\|v\|_2^2$, with $A<B$. Going through the proof we see that  using \eqref{expectgen} instead of \eqref{expecttight} from Lemma~\ref{lem:expect} we get
 \begin{align}
\E_y \left( \max_{|I|=S} \| P_{I}(\dico) y \|_2^2 \right)&= \E_{p, \sigma} \left(\max_{|I|=S} \| P_{I_p}(\dico) \dico \amp_{p, \sigma} \|_2^2 \right) \notag \\
&={K \choose S}^{-1} \left( \frac{1-\gamma_S^2}{K-S}\sum_I \| P_I(\dico) \dico\|_F^2 + \left(\frac{\gamma_S^2}{S}-\frac{1-\gamma_S^2}{K-S}\right)\sum_I \| P_I(\dico) \dico_I\|_F^2\right),\notag
\end{align} 
and
 \begin{align}
\E_y\left(\max_{|I|=S}   \| P_{I}(\pdico) y \|_2^2 \right)
&\geq \E_{p, \sigma} \left(\max_{|I|=S} \| P_{I_p}(\pdico) \dico \amp_{p, \sigma} \|_2^2 \right)\notag\\
&= {K \choose S}^{-1} \left( \frac{1-\gamma_S^2}{K-S}\sum_I \| P_I(\pdico) \dico\|_F^2 + \left(\frac{\gamma_S^2}{S}-\frac{1-\gamma_S^2}{K-S}\right)\sum_I \| P_I(\pdico) \dico_I\|_F^2\right).\notag
\end{align} 
Moreover by replacing $A$ with $B$ in \eqref{eq:worstcase} and \eqref{pessimistic} we get the new upper bound,
\begin{align}
\E_y\left(\max_{|I|=S}   \| P_{I}(\pdico) y \|_2^2 \right)
&\leq \E_{p, \sigma} \left(\max_{|I|=S} \| P_{I_p}(\pdico) \dico \amp_{p, \sigma} \|_2^2 \right)+ 4 B \eta_S\max_{|I|=S}  \|P_I(\pdico) - P_I(\dico)\|_{F}.
\end{align}
Since $B \eta_S$ is still of order $o(\eps^2)$ to prove that $\dico$ is a local maximum it suffices to show that up to second order $ \E_{p, \sigma} \left(\max_{|I|=S} \| P_{I_p}(\dico) \dico \amp_{p, \sigma} \|_2^2\right)-  \E_{p, \sigma} \left(\max_{|I|=S} \| P_{I_p}(\pdico) \dico \amp_{p, \sigma} \|_2^2\right)>0$. Conversely if we can find perturbation directions $z_i$ such that the reversed inequality holds, $\dico$ is not a local maximum. 
Using \eqref{pertproj} from the appendix, we get
\begin{align}
{K \choose S}&\left( \E_{p, \sigma} \left(\max_{|I|=S} \| P_{I_p}(\dico) \dico \amp_{p, \sigma} \|_2^2\right)-  \E_{p, \sigma} \left(\max_{|I|=S} \| P_{I_p}(\pdico) \dico \amp_{p, \sigma} \|_2^2\right)\right)\notag \\
&\quad =  \frac{1-\gamma_S^2}{K-S}\sum_I \left(\| P_I(\dico) \dico\|_F^2-\| P_I(\pdico) \dico\|_F^2 \right) +\lambda_S \sum_I \left(\| P_I(\dico) \dico_I\|_F^2 -\| P_I(\pdico) \dico_I\|_F^2 \right)\notag \\
&\quad =  \frac{1-\gamma_S^2}{K-S}\sum_I \tr \left(\dico^\star \left(P_I(\dico) - P_I(\pdico) \right) \dico\right) +\lambda_S \sum_I \| Q_I(\dico)B_I \|^2_F + O(\eps^3)\notag\\
&\quad = \frac{1-\gamma_S^2}{K-S}\sum_I 2 \tr \left(\dico^\star Q_I(\dico)B_I \dico_I^\dagger \dico \right) +O(\eps^2) +\lambda_S \sum_I \| Q_I(\dico)B_I \|^2_F + O(\eps^3)
\end{align}
The term $\sum_I \tr \left(\dico^\star Q_I(\dico)B_I \dico_I^\dagger \dico \right)$ is linear in $B$ and thus can be negative. Since it is also of order $O(\eps)$ whenever $\gamma_S < 1$ a necessary condition to have a local maximum exactly at $\dico$ is that for all $B_I=Z_I W_I A_I^{-1}$,
\begin{align}\label{nesscond}
\sum_I \tr \left(\dico^\star Q_I(\dico)B_I \dico_I^\dagger \dico \right) = 0.
\end{align}
In case $S=1$ we have $Q_i(\dico) b_i = b_i$ since $b_i \perp \atom_i$ and the condition above reduces to 
\begin{align}
\sum_i \tr \left(\dico^\star b_i \atom_i^\star \dico \right) =0
\quad \Leftrightarrow \quad \sum_i \atom_i^\star \dico \dico^\star b_i = 0
\end{align}
Choosing in turn $\omega_k = 0$ except for $k=i$ this means that for all $i$ and $z_i \perp \atom_i$ we need to have
\begin{align}
\frac{\omega_i}{\alpha_i}\ip{z_i}{\dico \dico^\star \atom_i}=0,
\end{align}
which is equivalent to every atom $\atom_i$ being an eigenvector of the frame operator, i.e. $\dico\dico^\star \atom_i=\lambda_i \atom_i, \, \forall i$. While this condition is certainly fulfilled when $\dico$ is a tight frame (corresponding to $\lambda_i = A$), it is sufficient for $\dico$ to be a collection of $m$ tight frames for $m$ orthogonal subspaces of $\R^\ddim$ - corresponding to the case $\dico=(\dico_{\lambda_1},\ldots ,\dico_{\lambda_m})$ with $\dico\dico^\star \dico_{\lambda_i}=\lambda_i \dico_{\lambda_i}$. Going through the same analysis as in the proof of Theorem~\ref{th:simplemodel} we see that in this second case $\dico$ is again a local maximum under the additional condition 
that $\amp_1^2> \frac{B-A+1}{B-A+K}$, where $A = \min_i \lambda_i$ and $B = \max_i \lambda_i$. \\
In case $S>1$, Condition \eqref{nesscond} is again implied by tightness of the dictionary but it is an open question whether conversely it implies tightness of the dictionary. However, for simplicity we will henceforth restrict our analysis to the situation where $\dico$ is a tight frame.
\end{remark}

\subsection{A continuous model of decaying coefficients}
After proving a recovery result for the simple coefficient model of the last section we would like to extend it to a wider range of coefficient distributions, especially continuous ones. \\
Looking back at the proof of Theorem~\ref{th:simplemodel} we see that apart from the condition ensuring optimality of the projection $P_{I_p}$ it also relied heavily on the equal probability of all sign sequences and permutations changing our base coefficient sequence. We therefore make the following definition.

\begin{definition} 
A probability measure $\nu$ on the unit sphere $S^{K-1} \subset \R^K$ is called symmetric if for all measurable sets $\mathcal{X}\subseteq S^{K-1}$, for all sign sequences $\sigma \in \{-1,1\}^K$ and all permutations $p$ we have
\begin{align}
\nu( \sigma \mathcal X)=\nu(\mathcal X), \quad &\mbox{where} \quad \sigma \mathcal X := \{ (\sigma_1 x_1, \ldots, \sigma_K x_K ) : x \in \mathcal{X} \}, \quad \mbox{and}\\
\nu( p( \mathcal X))=\nu(\mathcal X), \quad &\mbox{where} \quad p(\mathcal X ) := \{ ( x_{p(1)}, \ldots, x_{p(K)} ) : x \in \mathcal{X} \}.
\end{align}
\end{definition}

We are now ready to state a version of Theorem~\ref{th:simplemodel} for more general coefficient distributions.
\begin{theorem}\label{th:contmodel}
Let $\dico$ be a unit norm tight frame with frame constant $A=\natoms/\ddim$ and lower isometry constant $\delta_S$.
Let $x$ be drawn from a symmetric probability distribution $\nu$ on the unit sphere and assume that the signals are generated as $y=\dico x$. If there exists $\kappa>0$ such that for $c(x)$ a non-increasing rearrangement of the absolute values of $x$ and $I_p:=p^{-1}\left(\{1,\ldots \sparsity\}\right)$ we have, 
\begin{align}\label{maxatIpcont}
\nu\left(\min_{p,\sigma} \left(\|P_{I_p}(\dico)\dico \amp_{p,\sigma}(x)\|_2 - \max_{|I| = S,I\neq I_p}\|P_{I}(\dico)\dico \amp_{p,\sigma}(x)\|_2 \right)\geq 2\kappa\right) = 1
\end{align}
then there is a local maximum of \eqref{maxexpect} at $\dico$.\\
Moreover for $\pdico\neq \dico$ we have $\E_y\left(\max_{|I| = S} \| P_I(\pdico) y\|_2^2\right) < \E_y\left(\max_{|I| = S} \| P_I(\dico) y\|_2^2\right)$ as soon
as 
\begin{align}\label{eq:epsmaxcont}
d(\dico,\pdico) \leq  \frac{\kappa \sqrt{1-\delta_S} }{\sqrt{\frac{9S}{2}}\left(1+ 4\sqrt{\log\left(\frac{60 A {K \choose S}^2}{ \kappa \bar \lambda_S (1-\delta_S) }\right)} \right)},
\end{align}
where $\bar \lambda_S =  \frac{E_x\left(\amp_1^2(x)+\ldots+\amp_\sparsity^2(x)\right)}{S}-\frac{1-E_x\left(\amp_1^2(x)+\ldots+\amp_\sparsity^2(x)\right)}{K-S} $ and $\delta_S <1$ because of \eqref{maxatIpcont}.
\end{theorem}
\begin{proof} Let $\amp$ denote the mapping that assigns to each $x\in S^{K-1}$ the non increasing rearrangement of the absolute values of its components, \ie $\amp_i(x)=|x_{p(i)}|$ for a permutation $p$ such that $\amp_1(x)\geq \amp_2(x) \geq \ldots \geq \amp_K(x)\geq 0$. Then the mapping $\amp$ together with the probability measure $\nu$ on $S^{\natoms-1}$ induces a pull-back probability measure $\nu_\amp$ on $\amp(S^{\natoms-1})$, by $\nu_\amp (\Omega):=\nu(\amp^{-1}(\Omega))$ for any measurable set $\Omega\subseteq \amp(S^{\natoms-1})$. With the help of this new measure we can rewrite the expectations we need to calculate as,
\begin{align}
\E_y\left(\max_{|I| = S} \| P_I(\dico) y\|_2^2\right)& =
\E_x \left(\max_{|I| = S} \| P_I(\dico) \dico x \|_2^2 \right) \notag \\
&= \int_{x}  \max_{|I| = S} \| P_I(\dico) \dico x \|_2^2  d\nu(x)\notag\\
& = \int_{\amp(x)} \E_p \E_\sigma  \max_{|I| = S} \| P_I(\dico) \amp_{p,\sigma}(x)\|_2^2  d\nu_\amp(x).
\end{align}
The expectation inside the integral should seem familiar. Indeed we have calculated it already in the proof of Theorem~\ref{th:simplemodel} for $\amp(x)$ a fixed decaying sequence satisfying
\begin{align}
\|P_{I_p}(\dico)\dico \amp_{p,\sigma}\|_2 - \max_{|I| = S,I\neq I_p}\|P_{I}(\dico)\dico \amp_{p,\sigma}\|_2 \geq 2\kappa, \quad \forall \sigma, p.
\end{align}
By \eqref{maxatIpcont} this property is satisfied almost surely and so by applying Lemma~\ref{lem:expect} we get,
\begin{align*}
\E_x \left(\max_{|I| = S} \| P_I(\dico) \dico x \|_2^2 \right) &  =
 \int_{\amp(x)} \E_p \E_\sigma \| P_{I_p}(\dico) \amp_{p,\sigma}(x)\|_2^2  d\nu_\amp(x)\\
& =\int_{\amp(x)}  \frac{A(1-\bar \gamma_S^2(x))S}{(K-S)} + \left(\frac{\bar \gamma_S^2(x)}{S}-\frac{1-\bar\gamma_S^2(x)}{K-S}\right){K \choose S}^{-1}\sum_{I: |I|=S} \| \dico_I\|_F^2 d\nu_\amp(x),
\end{align*}
where $\gamma_S^2(x):=\amp_1^2(x)+\ldots+\amp_\sparsity^2(x)$.
Since for the integral term we simply have 
\begin{align}
\int_{\amp(x)} \gamma_S^2(x)d\nu_\amp(x) 
= \E_x\left(\max_{|I|= S} \|x_I\|_2^2 \right) = \bar{\gamma_S}^2,
\end{align} we arrive at the following estimate analogue to \eqref{eq:expatdico} 
\begin{align}
\E_x \left(\max_{|I| = S} \| P_I(\dico) \dico x \|_2^2 \right) 
& = \frac{A(1-\bar \gamma_S^2)S}{(K-S)} + \left(\frac{\bar \gamma_S^2}{S}-\frac{1-\bar\gamma_S^2}{K-S}\right){K \choose S}^{-1}\sum_{I: |I|=S} \| \dico_I\|_F^2.
\end{align}
Using the same argument we also get an estimate for the expectation of a perturbed dictionary analogue to \eqref{eq:expatpdico}, i.e.
\begin{align}
\E_x \left(\max_{|I| = S} \| P_I(\pdico) \dico x \|_2^2 \right) &\leq   
 \frac{A(1-\bar \gamma_S^2)S}{(K-S)} + \left( \frac{\bar \gamma_S^2}{S}-\frac{1-\bar \gamma_S^2}{K-S}\right) {K \choose S}^{-1}\sum_{I:|I|=S} \| P_I(\pdico) \dico_I\|_F^2\notag \\
 &\hspace{4cm} +  4 \eta_\sparsity A\max_{|I|=S}  \|P_I(\pdico) - P_I(\dico)\|_{F}\label{eq:expatpdico} .
 \end{align}
where
\begin{align}
\eta_S = 2 \sum_{I: P_I(\pdico) \neq P_I(\dico)} \exp \left(\frac{-\kappa^2}{32 \| P_I(\pdico) - P_I(\dico)\|^2_F}\right).
\end{align}
The rest of the proof simply consists of replacing $\gamma_S$ with $\bar\gamma_S$ in the proof of Theorem~\ref{th:simplemodel}.
\end{proof}

\begin{remark}
\noindent (a) Again the abstract condition in~\eqref{maxatIpcont} can be satisfied, i.e. we can find $\kappa>0$, if we have $S\mu <1/2$ and
\begin{align}\label{gendecaycond}
\nu\left(c_S(x) > \frac{1-S\mu}{1-2S\mu} c_{S+1}(x) +  \frac{4 \mu}{1-2S\mu} \sum_{i>\sparsity+1} |\amp_i(x)|\right)=1.
\end{align}
\noindent (b) Note that with the available tools it is also be possible to extend Theorem~\ref{th:contmodel} to signal models with coefficient distributions approaching the limit in~\eqref{maxatIpcont}, i.e. $\kappa=0$. 
However to keep the presentation concise we will not go into further details here but refer the interested reader to \cite{sc13sampta} or \cite{sc13arxiv} for the proof idea and some simple example distributions approaching the limit in the case of an orthonormal basis.
\end{remark}

\subsection{Bounded white noise}
With the tools used to prove the two noiseless identification results in the last two subsections it is also possible to analyse the case of (very small) bounded white noise. 
\begin{theorem}\label{th:contmodelnoise}
Let $\dico$ be a unit norm tight frame with frame constant $A=K/d$ and lower isometry constant $\delta_S$.
Assume that the signals $y$ are generated as
$
y=\dico x + \noise
$,
where $x$ is drawn from a symmetric decaying probability distribution $\nu$ on the unit sphere $S^{\natoms-1}$ and $\noise$ is a bounded random white noise vector, i.e. there exist two constants $\nsigma, \nmax$ such that  $\|\noise\|_2\leq \nmax$ almost surely, $\E (\noise)=0$ and $\E(\noise\noise^\star)=\nsigma^2 I$. If there exists $\kappa>0$ such that for $c(x)$ a non-increasing rearrangement of the absolute values of $x$ and $I_p:=p^{-1}\left(\{1,\ldots \sparsity\}\right)$ we have, 
\begin{align}\label{maxatIpnoise}
\nu\left(\min_{p,\sigma} \left(\|P_{I_p}(\dico)\dico \amp_{p,\sigma}(x)\|_2 - \max_{|I| = S,I\neq I_p}\|P_{I}(\dico)\dico \amp_{p,\sigma}(x)\|_2 \right)\geq 2\kappa + 2\nmax \right) = 1,
\end{align}
then there is a local maximum of \eqref{maxexpect} at $\dico$.\\
Moreover for $\pdico\neq \dico$ we have $\E_y\left(\max_{|I| = S} \| P_I(\pdico) y\|_2^2\right) < \E_y\left(\max_{|I| = S} \| P_I(\dico) y\|_2^2\right)$ as soon
as 
\begin{align}\label{eq:epsmaxnoise}
d(\dico,\pdico) \leq  \frac{\kappa \sqrt{1-\delta_S} }{\sqrt{\frac{9S}{2}}\left(1+ 4\sqrt{\log\left(\frac{60 \An {K \choose S}^2}{ \kappa \bar \lambda_S (1-\delta_S) }\right)} \right)},
\end{align}
where $\bar \lambda_S =  \frac{E_x\left(\amp_1^2(x)+\ldots+\amp_\sparsity^2(x)\right)}{S}-\frac{1-E_x\left(\amp_1^2(x)+\ldots+\amp_\sparsity^2(x)\right)}{K-S} $ and
$\An =(\sqrt{A} + \nmax)^2$. Again $\delta_S <1$ is implied by \eqref{maxatIpnoise}.

\end{theorem}
\begin{proof}
We streamline the proof, since it relies on the same ideas as those of Theorem~\ref{th:simplemodel} and Theorem~\ref{th:contmodel}. 
For a noisy signal $y=\dico x + \noise=\dico \amp_{p,\sigma}(x) + \noise$ the condition in \eqref{maxatIpnoise} guarantees that the maximal response for the original dictionary $\dico$ is taken at $I_p$ , since we have
\begin{align}
\| P_{I_p}(\dico) y\|_2 &\geq \|  P_{I_p}(\dico)\dico  \amp_{p,\sigma}(x) \|_2 + \|  P_I(\dico)r \|_2 \geq  \|  P_{I_p}(\dico)\dico  \amp_{p,\sigma}(x) \|_2 - \nmax,\\
\| P_I(\dico) y\|_2 & \leq \|  P_I(\dico)\dico \amp_{p,\sigma}(x)\|_2 + \|  P_I(\dico)r \|_2 \leq  \|  P_I(\dico)\dico \amp_{p,\sigma}(x)\|_2 + \nmax.
\end{align}
Thus we get,
\begin{align}\label{expdiconoise} \notag
\E_y\left(\max_{|I| = S} \| P_I(\dico) y\|_2^2\right)&
=\E_{r,x} \left(\max_{|I| = S} \| P_{I}(\dico) y\|_2^2\right)\notag\\
&= \E_r \left(\int_{\amp(x)} \E_p \E_\sigma  \max_{|I| = S} \| P_I(\dico) \left( (\dico) \amp_{p,\sigma}(x)+ r\right)\|_2^2  d\nu_\amp(x)\right)\notag\\
&= \E_r \left(\int_{\amp(x)} \E_p \E_\sigma  \| P_{I_p} \left( (\dico) \amp_{p,\sigma}(x)+ r\right) \|_2^2  d\nu_\amp(x)\right)\notag\\
&= \int_{\amp(x)} \E_p \E_\sigma \E_r \left( \| P_{I_p} (\dico) \left( (\dico) \amp_{p,\sigma}(x)+ r\right) \|_2^2 \right)  d\nu_\amp(x)\notag\\
&= \int_{\amp(x)} \E_p \E_\sigma \left( \| P_{I_p} (\dico) \amp_{p,\sigma}(x)\| + \E_r  \| P_{I_p}(\dico) r \|_2^2 \right) d\nu_\amp(x)\notag\\
&= \frac{A(1-\bar \gamma_S^2)S}{(K-S)} + \left(\frac{\bar \gamma_S^2}{S}-\frac{1-\bar\gamma_S^2}{K-S}\right){K \choose S}^{-1}\sum_{I: |I|=S} \| \dico_I\|_F^2+ S \nsigma^2.
\end{align}
For a perturbed dictionary and a noisy signal $y$ we can bound the response using the set $I$ analogue to \eqref{eq:worstcase},
\begin{align} 
 \| P_I(\pdico) y\|^2_2& \leq \| P_I(\dico) y\|^2_2 + 2 \| y \|^2_2  \| P_I(\pdico) - P_I(\dico)\|_{2,2} \notag \\
& \leq  \| P_I(\dico) y\|^2_2 +2(\sqrt{A} +\nmax)^2 \max_{|I| = S}\|P_I(\pdico) - P_I(\dico)\|_{F}\notag\\
& \leq  \| P_{I_p}(\dico) y\|^2_2 +2(\sqrt{A} +\nmax)^2 \max_{|I| = S}\|P_I(\pdico) - P_I(\dico)\|_{F}.
\end{align}
Reversing the roles of $\pdico$ and $\dico$ and setting $I=I_p$ in the inequality above then leads to 
\begin{align} 
 \| P_I(\pdico) y\|^2_2& \leq  \| P_{I_p}(\pdico) y\|^2_2 +4(\sqrt{A} +\nmax)^2 \max_{|I| = S}\|P_I(\pdico) - P_I(\dico)\|_{F}.
\end{align} 
Using the sets $\Sigma_p$ as defined in \eqref{defSigmap} we get the following estimate for $y=\dico \amp_{p,\sigma}(x)+ r$ with $\sigma \notin \Sigma_p$ and all $I \neq I_p$,
\begin{align} 
\| P_I (\pdico) y\|_2 & \leq \| P_I(\pdico)\amp_{p,\sigma}(x) \|_2 + \| P_I(\dico) r \|_2 \notag\\
&\leq \| P_I(\dico)\amp_{p,\sigma}(x) \|_2 + \| \left( P_I(\pdico) - P_I(\dico)\right) \amp_{p,\sigma}(x) \|_2 + \nmax \notag\\
&\leq \| P_I(\dico)\amp_{p,\sigma}(x) \|_2 + \kappa + \nmax \notag \\
&\leq \| P_{I_p}(\dico)\amp_{p,\sigma}(x) \|_2 - \kappa - \nmax \notag \\
&\leq \| P_{I_p}(\dico)\amp_{p,\sigma}(x) \|_2 - \| \left( P_{I_p}(\pdico) - P_{I_p}(\dico)\right) \amp_{p,\sigma}(x) \|_2 - \nmax \notag\\
& \leq \| P_{I_p}(\pdico)\amp_{p,\sigma}(x) \|_2 - \| P_{I_p}(\dico) r \|_2 \leq \| P_{I_p} (\pdico) y\|_2.
\end{align}
Thus we can estimate the expectation for a perturbed dictionary as
\begin{align}\label{exppdiconoise}
\E_y\left(\max_{|I| = S} \| P_I(\pdico) y\|_2^2\right)&
=\E_{r,x} \left(\max_{|I| = S} \| P_{I}(\pdico) y\|_2^2\right)\notag\\
&= \E_r \left(\int_{\amp(x)} \E_p \E_\sigma \left( \max_{|I| = S} \| P_I(\pdico) \left( \dico \amp_{p,\sigma}(x)+ r\right)\|_2^2\right) d\nu_\amp(x)\right)\notag\\
&= \E_r \left(\int_{\amp(x)} \E_p \left( \sum_{\sigma \in \Sigma_p} \max_{|I| = S}\| \ldots\|_2^2+ \sum_{\sigma \notin \Sigma_p} \max_{|I| = S}\| \ldots\|_2^2 \right)d\nu_\amp(x)\right)\notag\\
&\leq \E_r \left(\int_{\amp(x)} \E_p \E_\sigma \left( \| P_{I_p} (\pdico) \left( \dico \amp_{p,\sigma}(x)+ r\right) \|_2 \right) d\nu_\amp(x)\right)\notag\\
&\hspace{4cm} + 4\eta_S(\sqrt{A} +\nmax)^2 \max_{|I| = S}\|P_I(\pdico) - P_I(\dico)\|_{F}\notag\\
&\leq  \frac{A(1-\bar \gamma_S^2)S}{(K-S)} + \left(\frac{\bar \gamma_S^2}{S}-\frac{1-\bar\gamma_S^2}{K-S}\right){K \choose S}^{-1}\sum_{I: |I|=S} \| P_I(\pdico)\dico_I\|_F^2+ S \nsigma^2 \notag\\
&\hspace{4cm} + 4\eta_S(\sqrt{A} +\nmax)^2 \max_{|I| = S}\|P_I(\pdico) - P_I(\dico)\|_{F}.
\end{align}
The rest of the proof simply consists of replacing $\gamma_S$ with $\bar\gamma_S$ and $A$ with $(\sqrt{A} +\nmax)^2$ in the proof of Theorem~\ref{th:simplemodel}.
\end{proof}

\section{Finite sample size results}\label{sec:finite}
Finally we make the step from the asymptotic identification results derived in the last section to an identification result for a finite number of training samples. We consider the maximisation problem,
\begin{align}\label{maxcrit_finitesamp}
\max_{\pdico \in \mathcal{D}} \frac{1}{N} \sum_{n=1}^N \max_{|I| = S} \| P_I(\pdico) y_n\|_2^2.
\end{align}
The main idea is that whenever $\pdico$ is near to $\dico$ we have
\begin{align*}
\frac{1}{N} \sum_{n=1}^N \max_{|I| = S} \| P_I(\pdico) y_n\|_2^2 \approx \E_y \left( \max_{|I| = S} \| P_I(\pdico) y\|_2^2\right)
 <   \E_y \left( \max_{|I| = S} \| P_I(\dico) y\|_2^2\right) \approx \frac{1}{N} \sum_{n=1}^N \max_{|I| = S} \| P_I(\dico) y_n\|_2^2.
 \end{align*}
Concretising the sharpness of $\approx$ quantitatively and making sure that it is valid for all possible $\eps$-perturbations at the same time, leads to the following theorem.
\begin{theorem}\label{th:finitesampS}
Let $\dico$ be a unit norm tight frame with frame constant $A=K/d$ and lower isometry constant $\delta_S<1-\frac{S}{d}$.
Assume that the signals $y_n$ are generated as
$y_n=\dico x_n$,
where $x_n$ is drawn from a symmetric decaying probability distribution $\nu$ on the unit sphere $S^{\natoms-1}$ and $\noise$ is a bounded random white noise vector with $\|\noise\|_2\leq \nmax$ almost surely, $\E (\noise)=0$ and $\E(\noise\noise^\star)=\nsigma^2$. Further assume that there exists $\kappa>0$ such that for $c(x)$ a non-increasing rearrangement of the absolute values of $x$ and $I_p:=p^{-1}\left(\{1,\ldots \sparsity\}\right)$ we have, 
\begin{align}\label{maxatIpnoise2}
\nu\left(\min_{p,\sigma} \left(\|P_{I_p}(\dico)\dico \amp_{p,\sigma}(x)\|_2 - \max_{|I| = S,I\neq I_p}\|P_{I}(\dico)\dico \amp_{p,\sigma}(x)\|_2 \right)\geq 2\kappa + 2\nmax \right) = 1.
\end{align}
Abbreviate $\bar \lambda_S =  \frac{E_x\left(\amp_1^2(x)+\ldots+\amp_\sparsity^2(x)\right)}{S}-\frac{1-E_x\left(\amp_1^2(x)+\ldots+\amp_\sparsity^2(x)\right)}{K-S} $,
$\An =(\sqrt{A} + \nmax)^2$ and $C_S :=1 - \frac{S}{d(1-\delta_S)}$.\\
If for some $0<q<1/4$ the number of samples $\nsig$ satisfies
\begin{align}\label{samplesize}
2 N^{-q}+ N^{-2q} \leq \frac{\kappa \sqrt{1-\delta_S} }{\sqrt{\frac{9S}{2}}\left(1+ 4\sqrt{\log\left(\frac{135 \An   K {K \choose S}}{ \kappa \bar \lambda_S C_S S (1-\delta_S) }\right)} \right)},
\end{align}
then except with probability 
\begin{align} \label{successprob}
\exp\left( - \frac{N^{1-4q}\bar \lambda_S^2 S^2 C_S^2}{4K^2\An ^2} + Kd
\log\left( \frac{ N K\An }{2 \bar \lambda_S S C_S} \right)
 \right),
\end{align}
 there is a local maximum of \eqref{maxcrit_finitesamp} resp. local minimum of \eqref{ksvdcriterion} within distance
at most $2N^{-q}$ to $\dico$, i.e. for the local maximum $\tilde\pdico$ we have $\d(\tilde\pdico,\dico)\leq 2N^{-q}$.
\end{theorem}
\begin{proof}
Conceptually we need to show that for some $\eps_{\min}(N)< \eps_{\max}(N)$ and with probability $p(N)$ for all perturbations $\pdico$ with $\eps_{\min}(N) \leq d(\pdico,\dico) \leq \eps_{\max}(N)$ we have 
\begin{align}
\frac{1}{N} \sum_{n=1}^N \max_{|I| = S} \| P_I(\pdico) y_n\|_2^2<\frac{1}{N} \sum_{n=1}^N \max_{|I| = S} \| P_I(\dico) y_n\|_2^2 
\end{align}
To do this we need to add three ingredients to the asymptotic results of Theorem~\ref{th:contmodelnoise}, 1) that with high probability for a fixed dictionary $\pdico$ the sum of signal responses concentrates around its expectation, 2) a dense enough net for the space of all perturbations and 3) and a Lipschitz-type bound for the mapping
$\pdico \longrightarrow \max_{|I| = S} \| P_I(\pdico) y_n\|_2^2$. Then we can argue that an arbitrary perturbation will be close to a perturbation in the net, for which the sum concentrates around its expectation. This expectation is in turn is smaller than the expectation of the generating dictionary, around which the sum for the generating dictionary concentrates.\\
We start with the Lipschitz-type bound for the mapping $\pdico \longrightarrow \max_{|I| = S} \| P_I(\pdico) y_n\|_2^2$ on the set of perturbations with $d(\pdico,\dico)\leq \epsmax$. Analogue to \eqref{eq:worstcase} we have for any index set $I$ of size $S$,
\begin{align} 
 \| P_I(\pdico) y\|^2_2 &\leq  \| P_I(\ppdico) y\|^2_2 +2\An \max_{|I| = S}\|P_I(\pdico) - P_I(\dico)\|_{F} \notag \\
& \leq  \max_{|I|=S}\| P_I(\ppdico) y\|^2_2 +2\An \max_{|I| = S}\|P_I(\pdico) - P_I(\dico)\|_{F}.
\end{align}
Since this is true for all $I$ we further get that
\begin{align*} 
 \max_{|I|=S} \| P_I(\pdico) y\|^2_2 \leq \max_{|I|=S}\| P_I(\ppdico) y\|^2_2 +2\An \max_{|I| = S}\|P_I(\pdico) - P_I(\dico)\|_{F},
\end{align*}
and reversing the roles of $\pdico$ and $\ppdico$ leads to
\begin{align} 
\left| \max_{|I|=S} \| P_I(\pdico) y\|^2_2 - \max_{|I|=S}\| P_I(\ppdico) y\|^2_2 \right| \leq 2\An \max_{|I| = S}\|P_I(\pdico) - P_I(\dico)\|_{F}.
\end{align}
From Lemma~\ref{lem:projpdico} we know that
\begin{align*}
\|P_I(\pdico)- P_I(\ppdico ) \|^2_F \leq \frac{ 2S \frac{d(\pdico,\ppdico)^2}{1-d(\pdico,\ppdico)^2}  } {\|\pdico_I^\dagger\|_{2,2}^{-1} \left( \|\pdico_I^\dagger\|_{2,2}^{-1} -2 \sqrt{S} \frac{ d(\pdico,\ppdico)}{\sqrt{1-d(\pdico,\ppdico)^2} } \right) }.
\end{align*}
Now note that $\|\pdico_I^\dagger\|_{2,2}^{-1} $ is simply the minimal singular value of $\pdico_I$. Since we have $\delta_S < 1- S/d$ we get,
\begin{align}
\|\pdico_I^\dagger\|_{2,2}^{-1}=\sigma_{\min}(\pdico_I) =\sigma_{\min}(\dico_I A_I + Z_IW_I)&\geq \sigma_{\min}(\dico_I)\sigma_{\min}(A_I) -\sigma_{\max}( Z_IW_I) \notag \\
&\geq \sqrt{1-\delta_S}(1-\eps^2/2) - \sqrt{S}\eps.
\end{align}
The combination of the last three estimates, together with some simplifications, using the fact that both $\eps$ and $d(\pdico,\ppdico)$ will be smaller than $\epsmax \leq \frac{\sqrt{1-\delta_S}}{21\sqrt{S}}$, leads to the final bound,
\begin{align}
\left| \max_{|I| = S} \| P_I(\pdico) y_n \|_2^2-\max_{|I| = S} \| P_I(\ppdico) y_n \|_2^2 \right|\leq  
d(\pdico,\ppdico) \cdot \frac{C_L A_r \sqrt{S}}{ \sqrt{1-\delta_S}},
\end{align}
with $C_L=3.139$.
Next for $Y_n =  \max_{|I| = S} \| P_I(\pdico) y_n \|_2^2$ we have $Y_n\in [0, \An]$ and therefore by Hoeffding's inequality,
\begin{align*}
\P\left(\left | \frac{1}{N} \sum_{n=1}^N \max_{|I| = S} \| P_I(\pdico) y_n \|_2^2- \E ( \max_{|I| = S} \| P_I(\pdico) y_1 \|_2^2)\right|\geq t\right)
\leq e^{-Nt^2/\An^2}.
\end{align*}
The last ingredient is a $\delta$-net for all perturbations $\pdico$ with $d(\pdico,\pdico)\leq \epsmax$, i.e. a finite set of perturbations $\net$ such that for every $\pdico$ we can find $\ppdico \in \net$ with $d(\pdico,\ppdico)<\delta$. Remembering the parametrisation of all $\eps$-perturbations from the proof of Theorem~\ref{th:simplemodel} we see that the space we need to cover is the product of $K$ balls with radius $\epsmax$ in $\R^d$. Following for example the argument in Lemma 2 of \cite{ve10} we know that for the $m$-dimensional ball of radius $\epsmax$ we can find a $\delta$ net $\net_d$ with 
\begin{align*}
\sharp \net_d \leq \left (\epsmax + \frac{2\epsmax}{\delta}\right)^{d}.
\end{align*}
Thus for the product of $K$ balls in $\R^{d}$ we can construct a $\delta$-net $\net$ as the product of $K$ $\delta$-nets $\net_d$. Assuming that $\delta <1$ we then have,
\begin{align*}
\sharp \net \leq \left (\epsmax + \frac{2\epsmax}{\delta}\right)^{Kd} \leq \left (\frac{3\epsmax}{\delta}\right)^{Kd}.
\end{align*}
Using a union bound we can now estimate the probability that for all perturbations in the net the sum of responses concentrates around its expectation, as
\begin{align*}
\P\left(\exists \ppdico \in \net : \left| \frac{1}{N} \sum_{n=1}^N \max_{|I| = S} \| P_I(\pdico) y_n \|_2^2- \E ( \max_{|I| = S} \| P_I(\pdico) y_1 \|_2^2)\right|\geq t\right)
\leq  \left (\frac{3\epsmax}{\delta}\right)^{Kd} e^{-Nt^2/\An^2}.
\end{align*}
We can now turn to the triangle inequality argument. For a perturbation $\pdico$ with $d(\pdico,\dico)=\eps\leq\epsmax $ we can find $\ppdico \in \net$ with $d(\pdico,\ppdico)\leq \delta$ and $d(\ppdico,\dico)=\bar\eps$. We then have
\begin{align}
\frac{1}{N} \sum_{n=1}^N &\max_{|I|= S} \| P_I(\pdico) y_n \|_2^2 - \frac{1}{N} \sum_{n=1}^N \max_{|I| =S} \| P_I(\dico) y_n \|_2^2 \notag \\
&= \frac{1}{N} \sum_{n=1}^N \max_{|I|= S} \| P_I(\pdico) y_n \|_2^2 -\frac{1}{N} \sum_{n=1}^N \max_{|I|= S} \| P_I(\ppdico) y_n \|_2^2  \notag \\
&\hspace{2cm} + \frac{1}{N} \sum_{n=1}^N \max_{|I|= S} \| P_I(\ppdico) y_n \|_2^2  - \E\left( \max_{|I| = S} \| P_I(\ppdico) y_n \|_2^2\right) \notag \\
&\hspace{4cm} + \E\left( \max_{|I| = S} \| P_I(\ppdico) y_n \|_2^2\right) -  \E\left( \max_{|I| = S} \| P_I(\dico) y_n \|_2^2\right) \notag\\
&\hspace{6cm}+ \E\left( \max_{|I| = S} \| P_I(\dico) y_n \|_2^2\right)-\frac{1}{N} \sum_{n=1}^N \max_{|I|= S} \| P_I(\dico) y_n \|_2^2  \notag\\
&\leq \E\left( \max_{|I| = S} \| P_I(\ppdico) y_n \|_2^2\right) -  \E\left( \max_{|I| = S} \| P_I(\dico) y_n \|_2^2\right) + 2t + \delta \frac{C_L A_r \sqrt{S}}{ \sqrt{1-\delta_S}}.
\end{align}
Using the expression for the respective expectations for a noisy signal from \eqref{expdiconoise} and
 \eqref{exppdiconoise} and the abbreviation $\bar \lambda_S =   \frac{\bar \gamma_S^2}{S}-\frac{1-\bar\gamma_S^2}{K-S}$ we get,
\begin{align}
\frac{1}{N} \sum_{n=1}^N \max_{|I|= S}&\| P_I(\pdico) y_n \|_2^2 - \frac{1}{N} \sum_{n=1}^N \max_{|I| =S} \| P_I(\dico) y_n \|_2^2 \notag \\
&\leq  4 \An \max_{|I| = S}\|P_I(\ppdico) - P_I(\dico)\|_{F}  \sum_{I: \ldots} \exp \left(\frac{-\kappa^2}{32 \| P_I(\ppdico) - P_I(\dico)\|^2_F}\right)\notag \\
&\hspace{3cm}+ \bar \lambda_S {K \choose S}^{-1} \sum_{|I|=S} \left( \| P_I(\ppdico) \dico_I\|_F^2 -\| \dico_I\|_F^2\right) + 2t + \delta \frac{C_L A_r \sqrt{S}}{ \sqrt{1-\delta_S}} \notag \\
&\leq \frac{4 \An C_1}{\sqrt{1-\delta_S}}\max_{|I|=S} \| \bar B_{I}  \|_F  {K \choose S}  \exp \left(\frac{-\kappa^2 (1-\delta_S)}{32 C^2_1\max_{|I|=S} \|\bar B_{ I}  \|_F^2} \right) \notag \\
&\hspace{3cm}-  C_2 \bar \lambda_S { {K \choose S}^{-1}} \sum_{|I|=S} \| Q_I(\dico)\bar B_I  \|^2_F  + 2t + \delta \frac{C_L A_r \sqrt{S}}{ \sqrt{1-\delta_S}}, \label{finitediff}
\end{align}
where we have used Lemma~\ref{lem:projpdico} and that $\| Q_I(\dico)\bar B_I  \|_F \leq \| \bar B_I\|_F $. Using the condition on the isometry constant we now derive a (for all practical purposes) sharper lower bound than simply $\max_I\| Q_{I}(\dico)\bar B_{I}  \|_F^2$ for the sum in the equation above,
\begin{align}
 {K \choose S}^{-1} \sum_{I} \| Q_I(\dico) \bar{B}_I\|_F^2& =  {K \choose S}^{-1} \sum_{I}\left( \| \bar{B}_I\|_F^2-  \| P_I(\dico) \bar{B}_I\|_F^2\right) \notag \\
& ={K \choose S}^{-1} {K-1 \choose S-1} \| \bar{B}\|_F^2 -  {K \choose S}^{-1} \sum_{I}  \| (\dico^\dagger_I)^\star \dico_I^\star \bar{B}_I\|_F^2 \notag\\
&\geq \frac{S}{K}  \| \bar{B}\|_F^2 -  {K \choose S}^{-1}\sum_{I}  \| \dico^\dagger_I\|^2_{2,2} \| \dico_I^\star \bar{B}_I\|_F^2 \notag\\
&\geq \frac{S}{K}   \| \bar{B}\|_F^2 - \frac{1}{1-\delta_S} { K \choose S}^{-1} \sum_{I} \| \dico_I^\star \bar{B}_I\|_F^2 \notag\\
&\geq \frac{S}{K}  \| \bar{B}\|_F^2 - \frac{1}{1-\delta_S} { K \choose S}^{-1}   { K-2 \choose S-2}  \| \dico^\star \bar{B}\|_F^2 \notag\\
&\geq \frac{S}{K}   \left(1 - \frac{A}{1-\delta_S}\frac{S-1}{K-1} \right)\| \bar{B}\|_F^2. 
\end{align}
Using $A=K/d$ and denoting the index set for which $\|\bar B_I\|_F$ is maximal by $\bar I$ then leads to the bound
\begin{align}
 {K \choose S}^{-1} \sum_{I} \| Q_I(\dico) \bar{B}_I\|_F^2 \geq \frac{S}{K}   \left(1 - \frac{S}{d(1-\delta_S)} \right) \max_{|I|=S} \|\bar B_{ I}  \|_F^2.
\end{align}
Substituting the estimate above into \eqref{finitediff} we further get
\begin{align*}
\frac{1}{N} \sum_{n=1}^N &\max_{|I|= S} \| P_I(\pdico) y_n \|_2^2 - \frac{1}{N} \sum_{n=1}^N \max_{|I| =S} \| P_I(\dico) y_n \|_2^2\\
&\leq -  \frac{C_2 \bar \lambda_S  S}{K}   \left(1 - \frac{S}{d(1-\delta_S)} \right) \|\bar B_{\bar I}  \|_F^2\\
&\hspace{2cm}+ \frac{4 \An C_1}{\sqrt{1-\delta_S}} {K \choose S} \| \bar B_{\bar I}  \|_F    \exp \left(\frac{-\kappa^2 (1-\delta_S)}{32 C^2_1 \|\bar B_{\bar I}  \|_F^2} \right) + 2t + \delta \frac{C_L A_r \sqrt{S}}{ \sqrt{1-\delta_S}}, \notag 
\end{align*}
with $C_1=1.487$ and $C_2=0.897$.
Abbreviating $C_S= 1 - \frac{S}{d(1-\delta_S)} $ by Lemma~\ref{lem:trick} we have
\begin{align}
\frac{4 \An C_1}{\sqrt{1-\delta_S}} {K \choose S} \| \bar B_{\bar I}  \|_F    \exp \left(\frac{-\kappa^2 (1-\delta_S)}{32 C^2_1 \|\bar B_{\bar I}  \|_F^2} \right) \leq  \frac{(C_2-0.5) \bar \lambda_S C_S  S}{K}  \|\bar B_{\bar I}  \|_F^2
\end{align}
as soon as 
\begin{align}
\| B_{\bar I}  \|_F \leq \frac{4\kappa \sqrt{1-\delta_S} }{C_1\sqrt{32}\left(1+ 4\sqrt{\log\left(\frac{4\sqrt{32}C^2_1e^{1/16} AK {K \choose S}}{(C_2-0.5) \kappa\bar\lambda_S C_S S (1-\delta_S) }\right)} \right)},
\end{align}
which is satisfied if
\begin{align}
\bar{\eps} \leq  \frac{\kappa \sqrt{1-\delta_S} }{\sqrt{\frac{9S}{2}}\left(1+ 4\sqrt{\log\left(\frac{135 \An K {K \choose S}}{ \kappa \bar \lambda_S C_S S (1-\delta_S)  }\right)} \right)}:=\epsmax+\delta.
\end{align}
Under the condition above, which defines $\epsmax$ up to $\delta$, we further have
\begin{align}
\frac{1}{N} \sum_{n=1}^N \max_{|I| = S} \| P_I(\pdico) y_n \|_2^2 & - \frac{1}{N} \sum_{n=1}^N \max_{|I| = S} \| P_I(\dico) y_n \|_2^2 \notag \\
&\leq  - \frac{ \bar \lambda_S C_S S}{2K}  \bar\eps^2 + 2t + \delta \frac{C_L A_r \sqrt{S}}{ \sqrt{1-\delta_S}} \notag \\
&\leq  - \frac{ \bar \lambda_S C_S S}{2K} (\eps-\delta)^2 + 2t + \delta \frac{C_L A_r \sqrt{S}}{ \sqrt{1-\delta_S}}\notag \\
&\leq  - \frac{ \bar \lambda_S C_S S}{2K} \eps^2+ 2t + \delta ( \frac{(C_L+\epsmax) A_r \sqrt{S}}{ \sqrt{1-\delta_S}}.
\end{align}
We now choose $t=N^{-2q} \frac{ \bar \lambda_S C_S  S}{2K}   $ and $\delta= N^{-2q}  \frac{ \bar \lambda_S C_S \sqrt{S(1-\delta_S)}}{(C_L+\epsmax) \An K}  $ to get, that
except with probability,
\begin{align}
\exp\left( - \frac{N^{1-4q} \bar \lambda_S^2 C_S^2 S^2 }{4K^2\An^2} + Kd
\log\left( \frac{3  \epsmax (C_L+\epsmax) \An K N^{2q} }{\bar \lambda_S C_S \sqrt{S(1-\delta_S)} } \right)
 \right),
\end{align}
we have
\begin{align}
\frac{1}{N} \sum_{n=1}^N \max_{|I| = S} \| P_I(\pdico) y_n \|_2^2 - \frac{1}{N} \sum_{n=1}^N \max_{|I| = S} \| P_I(\dico) y_n \|_2^2&
\leq  - \frac{ \bar \lambda_S C_S S}{2K} ( \eps^2 - 4 N^{-2q}),
\end{align}
which is smaller than zero as long as $ \eps> 2N^{-q}:=\epsmin$. The statement follows from the bound $ 3 \epsmax(C_L+\epsmax)\leq \frac{\sqrt{1-\delta_S}}{2\sqrt{S}}$, and ensuring that $\epsmin < \epsmax$ using the crude bound $\delta \leq N^{-2q}$.
\end{proof}

\begin{remark}
Note that in case $S=1$ the above theorem is not only a result for the K-SVD minimisation principle but actually for K-SVD. While for $S>1$ the decay-condition is not strong enough to ensure that the sparse approximation algorithm used for K-SVD always finds the best approximation as soon as we are close enough to the generating dictionary, in the case $S=1$
any simple greedy algorithm, e.g. thresholding, will always find the best $1$-term approximation to any signal given any dictionary. Thus given the right initialisation and sufficiently many training samples K-SVD can recover the generating dictionary up to the prescribed precision with high probability. To make the theorem more applicable we quickly concretise how the distance between the generating dictionary $\dico$ and the local minimum output by K-SVD $\tilde{\pdico}$ decreases with the sample size. If we want the success probability to be of the order $1-N^{-Kd}$
we need
\begin{align*}
 \frac{-N^{1-4q}\lambda_S^2}{4K^2 C_L^2}+Kd\log(NKC_L/\lambda_S) \approx - Kd \log N,
\end{align*}
or $N^{1-4q} \approx K^3 d \log N$ meaning that $-q \approx -\frac{1}{4} + \frac{\log K}{\log N}$. Thus we have
\begin{align}
\log \left( d(\dico,\tilde{\pdico}) \right) = -q \log N \approx -\frac{\log N}{4} + \log K\notag
\end{align} 
or 
\begin{align}
d(\dico,\tilde{\pdico}) \approx K N^{-1/4}.
\end{align} 
\end{remark}
%
%
\noindent Let us now turn to a discussion of our results.

\section{Discussion}\label{sec:discussion}
We have shown that the minimisation principle underlying K-SVD~\eqref{ksvdcriterion} can identify a tight frame with arbitrary precision from signals generated from a wide class of decaying coefficients
distributions, provided that the training sample size is large enough. For the case $S=1$ in particular this means that K-SVD in combination with a greedy algorithm can recover the generating dictionary up to prescribed precision. To illustrate our results we conducted two experiments. \\
\subsection{Experiments}
\begin{figure}[thb]
\centering
  \includegraphics[width=8cm]{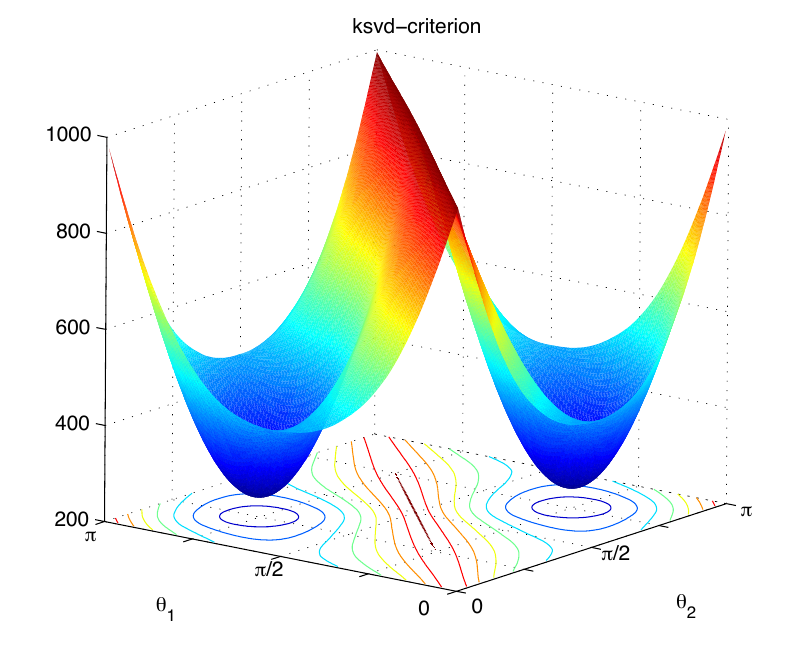}\\
    \caption{The K-SVD-criterion for the signals created from the decaying coefficients and an orthonormal basis, 
    the admissible dictionaries are parametrised by two angles $(\theta_1,\theta_2)$, i.e. $\atom_i = (\cos\theta_i, \sin\theta_i)$. \label{fig:ksvdcrit}}
\end{figure}

\begin{figure}[thb]
\begin{tabular}{cc}
 \includegraphics[width=8cm]{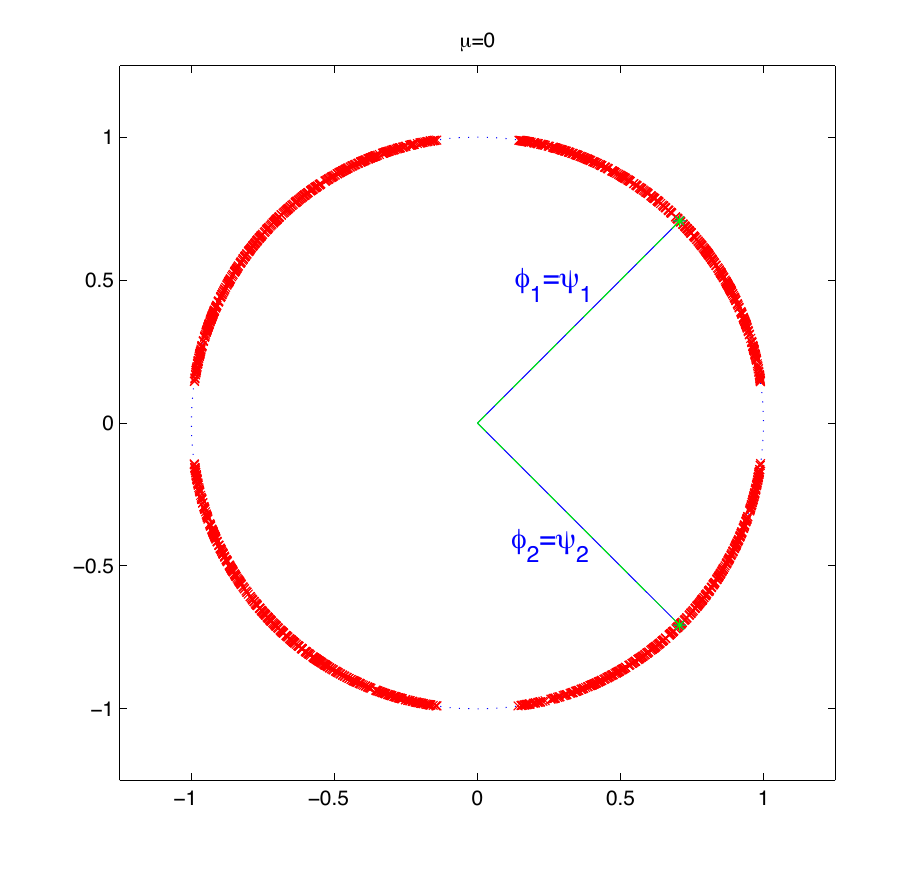} &
  \includegraphics[width=8cm]{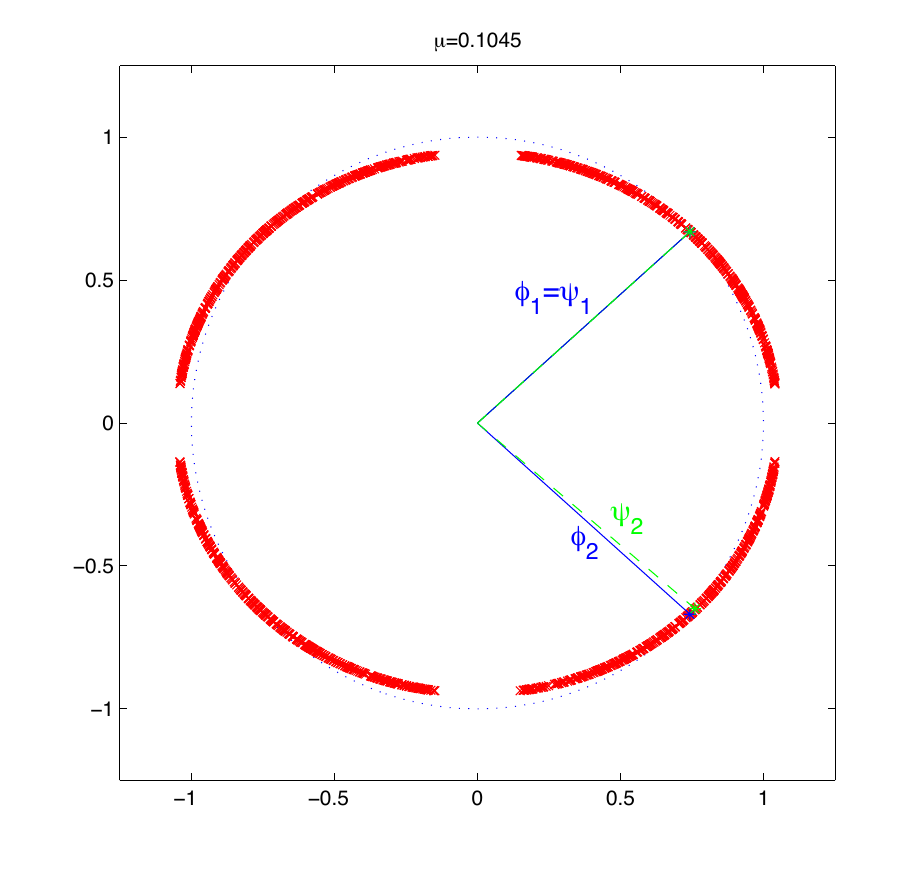} \\
   \includegraphics[width=8cm]{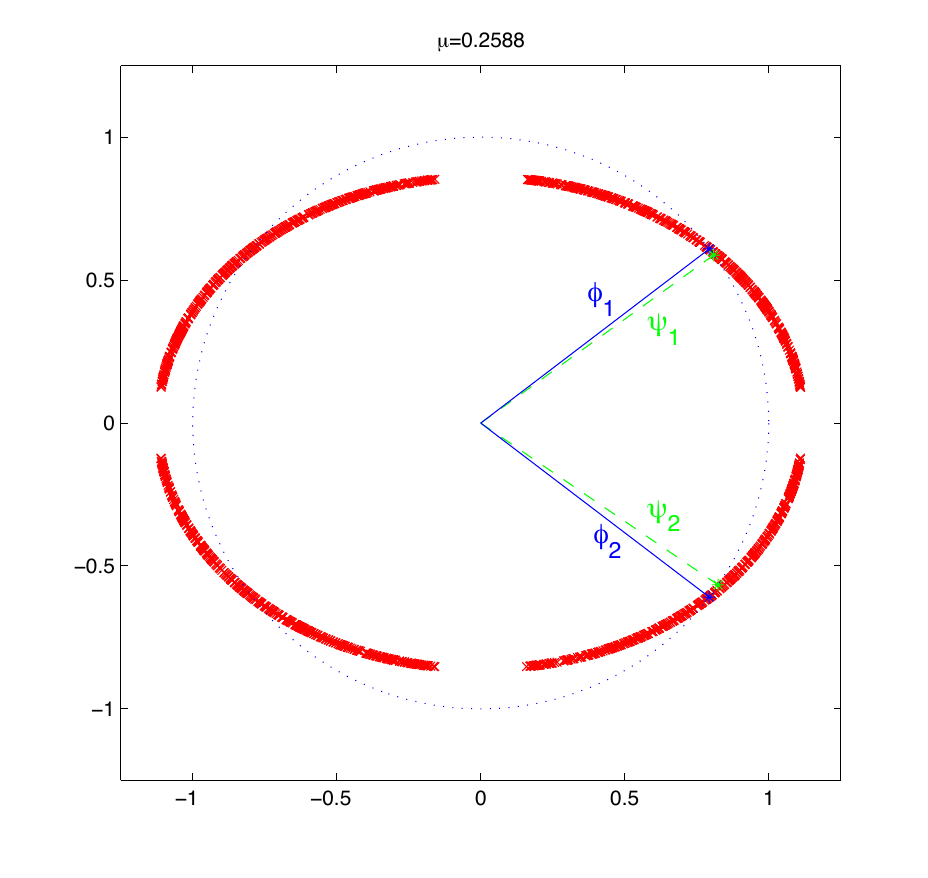} &
  \includegraphics[width=8cm]{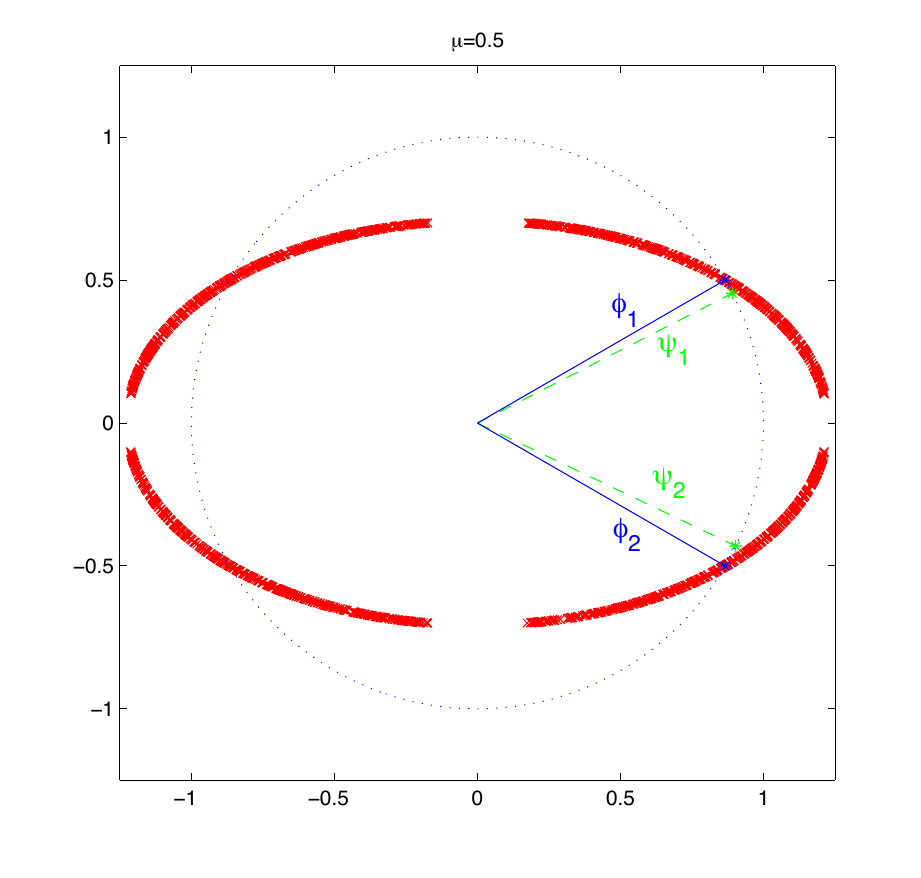} 
  \end{tabular}
    \caption{Signals created from various bases $\dico=(\atom_1,\atom_2)$ with increasing coherence $\mu$, together with the corresponding minimiser $\pdico=(\patom_1,\patom_2)$ of the 
    K-SVD-criterion for $\sparsity =1$.\label{fig:bases}}
\end{figure}

The first experiment demonstrates that the requirement on the dictionary to be tight in order to be identifiable translates to the case of finitely many training samples.
For simplicity and to allow for a visual representation of the outcome it was conducted in $\R^2$. We generated 1000 coefficients by drawing $\amp_2$ uniformly at random from the interval $[0,0.6]$, setting $c_1=\sqrt{1-c^2_2}$, randomly permuting the resulting vector and providing it with random $\pm$ signs. We then generated four sets of signals, using four bases with increasing coherence and the same coefficients, and for each set of signals found the minimiser of the K-SVD criterion~\eqref{ksvdcriterion} with $\sparsity=1$. Figure~\ref{fig:ksvdcrit} shows the objective function for the case of an orthonormal basis, while Figure~\ref{fig:bases} shows the four signal sets, the generating bases and the recovered bases. As predicted by our theoretical results when the generating basis is orthogonal it is also the minimiser of the K-SVD criterion, while for an oblique generating basis the minimiser is distorted towards the maximal eigenvector of the basis. Since for a 2-dimensional basis in combination with our coefficient distribution the abstract condition in~\eqref{maxatIpcont} is always fulfilled, this effect can only be due to the violation of the tightness-condition.\\

\begin{figure}[thb]
\centering
\begin{tabular}{cc}
 \includegraphics[width=8cm]{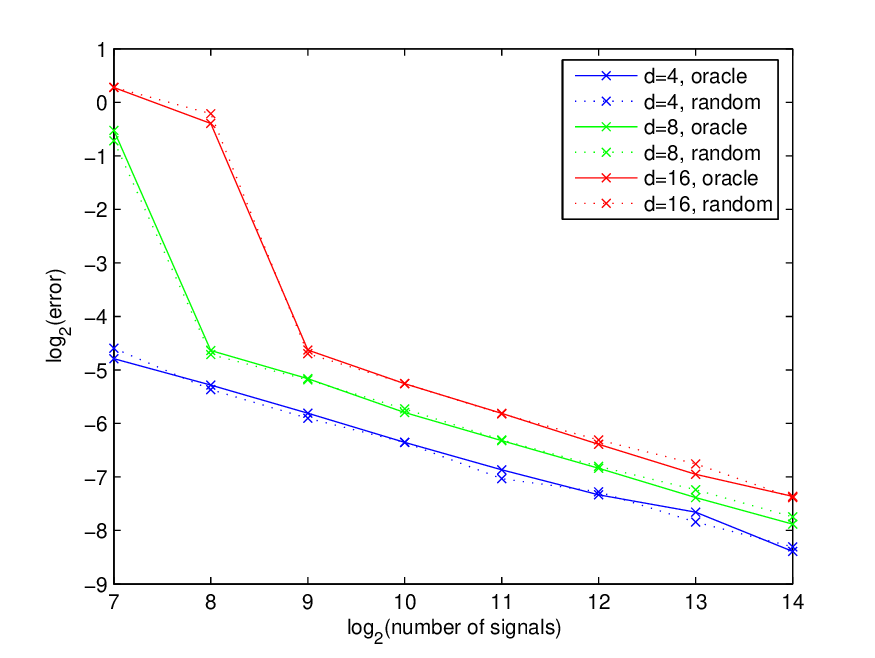} &
  \includegraphics[width=8cm]{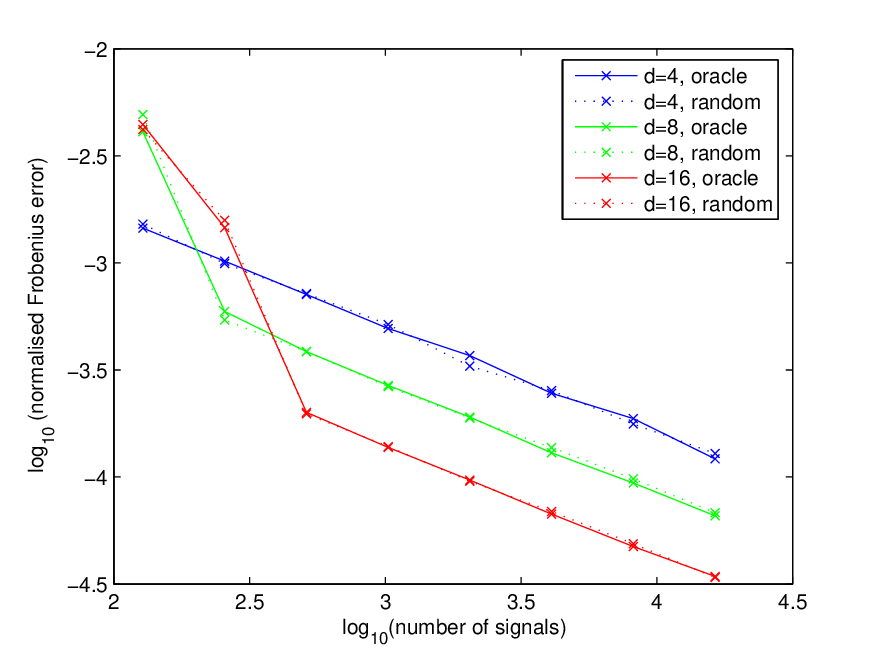}\\
  (a)&(b)
 \end{tabular}
    \caption{Error between the generating Hadamard-Dirac dictionary $\dico$ in $\R^d$ and the output $\tilde{\pdico}$ of the K-SVD algorithm with parameter $S=1$; the error is measured as $d(\dico,\tilde{\pdico})=\max_i \|\atom_i-\patom_i \|_2$) in (a) and as $\|\dico -\tilde{\pdico}\|_F/\sqrt{dK^3}$ in (b).\label{fig:errordecay}}
\end{figure}

The second experiment illustrates how the local minimum near the generating dictionary approaches the generating dictionary as the number of signals increases. As generating dictionary we choose the union of two orthonormal bases, the Hadamard and the Dirac basis, in dimension $d=4,8,16$, i.e. $K=2d$. We then generated $2$-sparse signals by first drawing $\amp_1$ uniformly at random from the interval $[0.99,1]$, setting $c_2=\sqrt{1-c_1^2}$, meaning $c_2 \in [0, 0.1]$, and $c_i = 0$ for $i\geq 3$ and then setting $y = \dico c_{\sigma, p}$ for a uniformly at random chosen sign sequence $\sigma$ and permutation $p$. We then run the original K-SVD algorithm as described in \cite{ahelbr06}, with a greedy algorithm, and sparsity parameter $S=1$, using both an oracle initialisation (i.e. the generating dictionary) and a random initialisation, on training sets containing $128 \cdot 2^n$ signals for n increasing from 0 to 7. Figure~\ref{fig:errordecay} (a) plots the maximal distance between two corresponding atoms of the generating and the learned dictionary, $d(\dico,\tilde{\pdico})=\max_i \|\atom_i-\patom_i \|_2$, averaged over 10 runs. Figure~\ref{fig:errordecay} (b) is designed to be comparable to the experiment conducted for the noisy $\ell_1$-criterion in \cite{bagrje13} and plots the normalised Frobenius norm between the generating and the learned dictionary, $\|\dico -\tilde{\pdico}\|_F/\sqrt{dK^3}$, averaged over 10 runs. \\

As expected we have a log-linear relation between the number of samples and the reconstruction error. However our predictions seem to be too pessimistic. So rather than an inclination of $-\frac{1}{4}$ we see one of $-\frac{1}{2}$ indicating that $d(\dico,\tilde{\pdico})\approx N^{-\frac{1}{2}}$. We also see that
both the oracle and the random initialisation lead to the same results, raising the question of uniqueness of the equivalent local minima, compare also \cite{bagrje13}.

\subsection{Future work}
Finally let us point out further research directions based on a comparison of our 
results for the K-SVD-minimisation principle to existing identification results.
Compared to the available identification results for the $\ell_1$-minimisation principle,
\begin{align}
\min_{\dico \in \mathcal D, X : Y=\dico X} \sum_{ij} |X_{ij}|,
\end{align} 
it seems at first glance that the
K-SVD-criterion requires a larger sample size than the $\ell_1$-criterion, i.e. $N^{1-4q}/\log N = O (K^3d)$ as opposed to $O(d^2 \log d)$ reported in \cite{grsc10} for a basis and $O(K^3)$ reported in \cite{gewawrXX} for an overcomplete dictionary. Also it does not allow for exact identification with high probability but only guarantees stability. However, this effect may be due to the more general signal model which assumes decay rather than exact sparsity. Indeed it is very interesting to compare our results to a recent result for a noisy version of the $\ell_1$-minimisation principle, \cite{bagrje13}, which provides stability results under unbounded white noise and, omitting log factors, also derives a sampling complexity of $O (K^3d)$.\\
Another difference, apparently intrinsic to the two minimisation criteria is that probably the K-SVD criterion can only identify tight dictionary frames exactly, while the $\ell_1$-criterion allows identification of arbitrary dictionaries. Thus to support the use of the K-SVD criterion for the learning of non-tight dictionaries also theoretically, we plan to study the stability of the K-SVD criterion under non-tightness by analysing the maximal distance between an original, non tight dictionary with condition number $\sqrt{B/A}>1$ and the closest local maximum, cp. also Figure~\ref{fig:bases}. \\
Compared to identification results for the ER-SpUD algorithm, \cite{spwawr12}, our results have the advantage of being valid also for overcomplete dictionaries and not exactly sparse signals. The disadvantage is that our results are valid only locally and in case $S>1$ only for a criterion, not an algorithm. An important research direction therefore is to analyse how close the output of K-SVD is to the local minimum of the K-SVD criterion given the same initialisation in the general case.\\
The last research direction we want to point out is how much decay of the coefficients is actually necessary. For the case $S=1$, it is quite easy to see, compare also \cite{sc13sampta, sc13arxiv}, that a condition of the type $\amp_1>\amp_{2} + 2\mu \|\amp\|_1$ ensures that the maximal inner product is always attained at $i_p=p^{-1}(1)$. However, typically we have $|\ip{\atom_{i}}{ \dico \amp_{p,\sigma}}|\approx  \amp_{p(i)} \pm\mu$. Therefore a condition such as $\amp_1>\amp_{2} + O(\mu)$, which allows for outliers, i.e. signals for which the maximal projection is not at $i_p$, might be sufficient to prove - if not exact identifiability - at least stability. Together with the inspiring techniques from \cite{bagrje13}, we expect the tools developed in the course of such an analysis to allow us also to deal with unbounded white noise.

\appendices

\section*{Acknowledgments}
This work was supported by the Austrian Science Fund (FWF) under Grant no. Y432 and J3335 and improved thanks to the reviewers' detailed and pointed comments.\\
Also I would like to thank Massimo Fornasier for SUPPORT (in capital letters), 
Maria Mateescu for proof-reading the proposal leading to grant J3335 and helping me with the shoeshine,
Remi Gribonval for pointing out the connection between an early version of \eqref{maxexpect} with $S=1$ and K-SVD and 
 Jan Vybiral for reading several ugly draft versions. 

\section{Technical details for the proof of Theorem~\ref{th:simplemodel}}
\begin{lemma}\label{lem:expect} For two frames $\dico,\pdico$ we have
 \begin{align}\label{expectgen}
\E_p \E_\sigma& \left(  \| P_{I_p}(\pdico) \dico\amp_{p,\sigma}\|_2^2 \right)\notag\\
&={K \choose S}^{-1} \left( \frac{1-\gamma_S^2}{K-S}\sum_I \| P_I(\pdico) \dico\|_F^2 + \left(\frac{\gamma_S^2}{S}-\frac{1-\gamma_S^2}{K-S}\right)\sum_I \| P_I(\pdico) \dico_I\|_F^2\right),
\end{align} 
where $\gamma_S^2:=\amp_1^2+\ldots+\amp_\sparsity^2$.\\
In case $\dico$ is a tight frame with frame constant A and $\delta_S(\pdico)<1$ this reduces to
 \begin{align}\label{expecttight}
\E_p \E_\sigma \left(  \| P_{I_p}(\pdico) \dico\amp_{p,\sigma}\|_2^2 \right)=
 \frac{A(1-\gamma_S^2)S}{K-S} + \left(\frac{\gamma_S^2}{S}-\frac{1-\gamma_S^2}{K-S}\right){K \choose S}^{-1}\sum_I \| P_I(\pdico) \dico_I\|_F^2.
\end{align}
\end{lemma}
\begin{proof}
We have
\begin{align}
\E_p \E_\sigma \left(  \| P_{I_p}(\pdico) \dico\amp_{p,\sigma}\|_2^2 \right)
&=\sum_{i} \E_p  \left(\amp^2_{p(i)}   \| P_{I_p}(\pdico) \atom_i\|_2^2  \right) \label{eq:sumexp2}
\end{align} 
For each $i$ we now split the set of all permutations $\mathcal{P}$ into disjoint sets $\mathcal{P}^i_{Ik}$, defined as
$$
\mathcal{P}^i_{Ik}:=\{p:p(I)=\{1,\ldots, \sparsity\}, p(i)=k\},
$$
where $I$ is a subset of $\{1,\ldots ,K\}$ with $|I|=\sparsity$ and $k=1\ldots \natoms$.
We then have $\mathcal{P}=\cup_{I,k}\mathcal{P}^i_{jk}$ and
$$
|\mathcal{P}^i_{Ik}|=\left\{
\begin{array}{ll}
  (K-S-1)!S!& \mbox{ if } i\notin I  \mbox{ and } k\geq \sparsity+1 \\
 (K-S)!(S-1)!&  \mbox{ if } i=j\in I \mbox{ and } k=p(j)\\
0& \mbox{ else}\\
 \end{array} \right..
 $$
 Using these sets we can compute the expectations in \eqref{eq:sumexp2} as follows 
\begin{align*}
\E_p   \left(\amp^2_{p(i)}   \| P_{I_p}(\pdico) \atom_i\|_2^2  \right)
&=\frac{1}{K!} \sum_I \sum_k \sum_{p\in \mathcal{P}^i_{Ik}} \amp^2_k \| P_I(\pdico) \atom_i\|_2^2 \\
&= {K \choose S}^{-1}\frac{1}{K-S} \sum_{I:i\notin I} \sum_{k\geq\sparsity+1} \amp^2_k \| P_I(\pdico) \atom_i\|_2^2 +{K \choose S}^{-1}\frac{1}{S}\sum_{I:i\in I} \sum_{k\leq\sparsity} \amp^2_k \| P_I(\pdico) \atom_i\|_2^2\\
&=  {K \choose S}^{-1}\left( \frac{1-\amp_1^2-\ldots-\amp_\sparsity^2}{K-S}\sum_{I:i\notin I} \| P_I(\pdico) \atom_i\|_2^2 + \frac{\amp_1^2+\ldots+\amp_\sparsity^2}{S}\sum_{I:i\in I} \| P_I(\pdico) \atom_i\|_2^2\right).
\end{align*}
Abbreviating $\gamma_S^2:=\amp_1^2+\ldots+\amp_\sparsity^2$ and re-substituting the above expression into \eqref{eq:sumexp2} leads to,
 \begin{align*}
{K \choose S}\E_p \E_\sigma \left(  \| P_{I_p}(\pdico) \dico\amp_{p,\sigma}\|_2^2 \right)&=
\frac{1-\gamma_S^2}{K-S}\sum_i\sum_{I:i\notin I} \| P_I(\pdico) \atom_i\|_2^2 + \frac{\gamma_S^2}{S}\sum_i\sum_{I:i\in I} \| P_I(\pdico) \atom_i\|_2^2\\
&=
 \frac{1-\gamma_S^2}{K-S}\sum_i\sum_{I} \| P_I(\pdico) \atom_i\|_2^2 + \left(\frac{\gamma_S^2}{S}-\frac{1-\gamma_S^2}{K-S}\right)\sum_i\sum_{I:i\in I} \| P_I(\pdico) \atom_i\|_2^2\\
&=
\frac{1-\gamma_S^2}{K-S}\sum_I\sum_{i} \| P_I(\pdico) \atom_i\|_2^2 + \left(\frac{\gamma_S^2}{S}-\frac{1-\gamma_S^2}{K-S}\right)\sum_I\sum_{i\in I} \| P_I(\pdico) \atom_i\|_2^2\\
&=
 \frac{1-\gamma_S^2}{K-S}\sum_I \| P_I(\pdico) \dico\|_F^2 + \left(\frac{\gamma_S^2}{S}-\frac{1-\gamma_S^2}{K-S}\right)\sum_I \| P_I(\pdico) \dico_I\|_F^2.
\end{align*} 
If $\dico$ is a tight frame and $\delta_S(\pdico)<1$, meaning $\pdico_I$ always has full rank, we have $\| P_I(\pdico) \dico\|_F^2=\tr(\dico^\star P_I(\pdico)^\star P_I(\pdico) \dico)=\tr(P_I(\pdico) \dico \dico^\star )=A\sparsity$, which leads to the second statement.
\end{proof}

\begin{lemma}\label{lem:projpdico}
Let $\dico$ be a dictionary with isometry constant $\delta_S <1$ and $\pdico$ be an $\eps$ perturbation of $\dico$, i.e. $d(\dico, \pdico)=\eps$. So we can write $\pdico=\dico A + Z W$, where $A=\diag((1-\eps_i^2/2)_i)$, $W=\diag((\eps_i^2-\eps_i^4/4)^{1/2})_i)$ for $\max_i \eps_i =\eps$ and $Z=(z_1, \ldots z_K)$ where $\ip{z_i}{\atom_i}=0$. Abbreviating $Q_I(\dico)=\I_\ddim-P_I(\dico)$, where $\I_d$ is the identity matrix in $\R^{d \times d}$, and $B_I=Z_I W_I A_I^{-1}$, where $A_I=\diag((1-\eps_i^2/2)_{i\in I})$ and $W_I=\diag((\eps_i^2-\eps_i^4/4)^{1/2})_{i\in I})$, we have 
\begin{align*}
\| P_I(\dico)-P_I(\pdico)\|^2_F\leq \frac{2\| Q_I(\dico)B_I \|^2_F}{\sqrt{1-\delta_S} \left(\sqrt{1-\delta_S} - 2\|B_I \|_{F}\right) },
\end{align*}
and
\begin{align*}
\| P_I(\pdico) \dico_I\|_F^2\leq \|\dico_I\|^2_F - \| Q_I(\dico)B_I \|^2_F + \frac{ 2 \| Q_I(\dico)B_I \|^2_F \|B_I \|_F}{\sqrt{1-\delta_S} - \|B_I \|_{F}}+\frac{\| Q_I(\dico)B_I \|^4_F}{\left( \sqrt{1-\delta_S}  - 2\|B_I \|_{F}\right)^2 },
\end{align*}
whenever $\eps$ is small enough.\\
In particular when $\eps \leq \frac{\sqrt{1-\delta_S}}{21\sqrt{S}}$ we have
\begin{align*}
\| P_I(\dico)-P_I(\pdico)\|_F\leq 1.487\cdot\frac{ \| Q_I(\dico)B_I \|_F}{\sqrt{1-\delta_S}}
\qquad \mbox{and} \qquad
\| P_I(\pdico) \dico_I\|_F^2\leq \|\dico_I\|^2_F - 0.897\cdot \| Q_I(\dico)B_I \|^2_F.
\end{align*}
\end{lemma}
\begin{proof}
We first compute the projection $P_I(\pdico)=\pdico_I (\pdico_I^\star \pdico_I)^{-1}\pdico_I^\star$  in terms of $\dico_I$ and $B_I$. Since $\delta_S <1$ the matrix $\dico^\star_I\dico_I$ is invertible and we can write $\dico_I^\dagger=(\dico^\star_I\dico_I)^{-1}\dico_I^\star$. We now split $\pdico_I$ into the part contained in the span of $\dico_I$ and the rest,
\begin{align}
\pdico_I&=P_I(\dico)\pdico_I+Q_I(\dico)\pdico_I \notag \\
&=\dico_I A_I + P_I(\dico)Z_I W_I+Q_I(\dico)Z_I W_I \notag \\
&=\left( \dico_I (\I_S + \dico_I^\dagger B_I) +Q_I(\dico)B_I \right) A_I. \label{eq:pdicodec}
\end{align}
Next we calculate $(\pdico_I^\star \pdico_I)^{-1}$. Using the expression in~\eqref{eq:pdicodec} we have
\begin{align*}
\pdico_I^\star \pdico_I&=A_I \left((\I_S + \dico_I^\dagger B_I)^\star\dico_I^\star \dico_I (\I_S + \dico_I^\dagger B_I) +B_I^\star Q_I(\dico)B_I \right) A_I.
\end{align*}
Using the fact that $\|\dico_I^\dagger\|_{2,2}^2=\|(\dico_I^\star \dico_I)^{-1}\|_{2,2}\leq (1-\delta_S)^{-1}$ we can estimate 
\begin{align}
\| \dico_I^\dagger B_I\|_{2,2}  \leq \| \dico_I^\dagger \|_{2,2}  \|B_I\|_F \leq \sqrt{1-\delta_S}\frac{ \eps\sqrt{S}}{\sqrt{1-\eps^2} }.
\end{align}
Since this is smaller than 1 for $\eps$ small enough, we can calculate the inverse of $(\I_S + \dico_I^\dagger B_I)$ using a Neumann series, i.e.
\begin{align*}
(\I_S + \dico_I^\dagger B_I)^{-1}= \I_S +\sum_{i=1}^\infty (- \dico_I^\dagger B_I)^i,
\end{align*}
with $\|(\I_S + \dico_I^\dagger B_I)^{-1}\|_{2,2} \leq (1-\| \dico_I^\dagger B_I\|_{2,2})^{-1}$.
This allows us to rewrite $\pdico_I^\star \pdico_I$ as,
\begin{align}
\pdico_I^\star \pdico_I&=A_I (\I_S + \dico_I^\dagger B_I)^\star\dico_I^\star \dico_I\left( \I_S + R_I \right)(\I_S + \dico_I^\dagger B_I) A_I,\notag \\
&\mbox{for }  R_I=(\dico_I^\star \dico_I)^{-1}(\I_S + \dico_I^\dagger B_I)^{\star-1} B_I^\star Q_I(\dico)B_I (\I_S + \dico_I^\dagger B_I)^{-1},\label{Rdef}
\end{align}
and we can estimate
\begin{align}\label{Rbound}
\|R_I\|_{2,2}&\leq \|(\dico_I^\star \dico_I)^{-1}\|_{2,2}\|(\I_S + \dico_I^\dagger B_I)^{-1}\|_{2,2}^2 \|Q_I(\dico)B_I \|_{2,2}^2\notag \\
&\leq\frac{ \|Q_I(\dico)B_I \|_{F}^2}{\left( \|\dico_I^\dagger\|_{2,2}^{-1} - \|B_I \|_{F}\right)^2}\\
& \leq \frac{ \frac{S\eps^2}{1-\eps^2}}{ 1-\delta_S - 2 \frac{S\eps^2}{1-\eps^2} }.\notag
\end{align}
For $\eps$ small enough this is again smaller than $1$ and so we can again use a Neumann series to calculate the inverse,
\begin{align}
(\pdico_I^\star \pdico_I)^{-1}=A_I^{-1}(\I_S+\dico_I^\dagger B_I)^{-1}\left(\I_S +\sum_{i=1}^\infty (-R_I)^i\right)(\dico_I^\star \dico_I)^{-1}(\I_S + \dico_I^\dagger B_I)^{-1\star}A_I^{-1}.\notag
\end{align}
Thus we finally get for the projection onto the perturbed atoms indexed by $I$,
\begin{align}\label{pertproj}
P_I(\pdico)&=\left( \dico_I+ Q_I(\dico)B_I (\I_S+\dico_I^\dagger B_I)^{-1}\right)\notag\\
& \hspace{2cm} \cdot \left(\I_S +\sum_{i=1}^\infty (-R_I)^i\right)(\dico_I^\star \dico_I)^{-1}\left( \dico_I+ Q_I(\dico) B_I (\I_S+\dico_I^\dagger B_I)^{-1}\right)^\star.
\end{align}
To calculate $\| P_I(\dico)-P_I(\pdico)\|^2_F$ up to order $O(\eps^2)$ we need to keep track of all terms involving $B_I$ up to second order. We have,
\begin{align}
\| P_I(\dico)-P_I(\pdico)\|^2_F&=\tr( P_I(\dico)) - \tr(P_I(\dico) P_I(\pdico)) + \tr(P_I(\pdico))\notag \\
&=2\sparsity - 2 \tr((\dico_I^\star \dico_I)^{-1}\dico_I^\star\pdico_I (\pdico_I^\star \pdico_I)^{-1}\pdico_I^\star \dico_I )\notag \\
&=2\sparsity - 2\tr \left(\I_S +\sum_{i=1}^\infty (-R_I)^i\right) \leq 2\sum_{i=1}^\infty \|R_I\|_F^i,
\end{align}
and employing the bound for $\|R_I\|_F$ from \eqref{Rbound} leads us to, 
\begin{align}
\| P_I(\dico)-P_I(\pdico)\|^2_F&\leq \frac{2\| Q_I(\dico)B_I \|^2_F}{\left( \|\dico_I^\dagger\|_{2,2}^{-1} - \|B_I \|_{F}\right)^2-\| Q_I(\dico)B_I \|^2_F}\notag\\
&\leq \frac{2\| Q_I(\dico)B_I \|^2_F}{\|\dico_I^\dagger\|_{2,2}^{-1}\left(\|\dico_I^\dagger\|_{2,2}^{-1}  - 2\|B_I \|_{F}\right) }\leq \frac{2\| Q_I(\dico)B_I \|^2_F}{\sqrt{1-\delta_S} \left(\sqrt{1-\delta_S} - 2\|B_I \|_{F}\right) }.
\end{align}
Similarily we get for $\| P_I(\pdico) \dico_I\|_F^2$,
\begin{align*}
\| P_I(\pdico) \dico_I\|_F^2&=\tr(\dico_I^\star \pdico_I (\pdico_I^\star \pdico_I)^{-1}\pdico_I^\star\dico_I)\\
&=\tr\left(\dico_I^\star \dico_I \left(\I_S +\sum_{i=1}^\infty (-R_I)^i\right)\right) \\
&=\tr\left(\dico_I^\star \dico_I\right) - \tr \left( \left(\I_S +\sum_{i=1}^\infty (- \dico_I^\dagger B_I)^i\right)^{\star} B_I^\star Q_I(\dico)B_I  \left(\I_S +\sum_{i=1}^\infty (- \dico_I^\dagger B_I)^i\right) \right)\\
&\hspace{4cm}  +  \tr \left(\dico_I^\star \dico_I \sum_{i=2}^\infty (-R_I)^i\right)\\
&= \tr\left(\dico_I^\star \dico_I\right) - \tr \left( B_I^\star Q_I(\dico)B_I \right) - 2 \tr \left( B_I^\star Q_I(\dico)B_I \sum_{i=1}^\infty (- \dico_I^\dagger B_I)^i \right)\\
&\hspace{1cm}- \tr \left(\left( \sum_{i=1}^\infty (- \dico_I^\dagger B_I)^i \right)^\star B_I^\star Q_I(\dico)B_I \sum_{i=1}^\infty (- \dico_I^\dagger B_I)^i \right)  +  \tr \left(\dico_I^\star \dico_I \sum_{i=2}^\infty (-R_I)^i\right).
\end{align*}
Taking into account that the fourth term in the above equation is always smaller than zero we finally get the bound,
\begin{align}
\| P_I(\pdico) \dico_I\|_F^2&\leq \|\dico_I\|^2_F - \| Q_I(\dico)B_I \|^2_F + 2 \| Q_I(\dico)B_I \|^2_F \sum_{i=1}^\infty \| \dico_I^\dagger B_I\|_F^i+ \|\dico_I^\star \dico_IR_I\|_F \sum_{i=1}^\infty \| R_I\|_F^i \notag \\
&\leq \|\dico_I\|^2_F - \| Q_I(\dico)B_I \|^2_F + \frac{ 2 \| Q_I(\dico)B_I \|^2_F \|B_I \|_F}{\|\dico_I^\dagger\|_{2,2}^{-1} - \|B_I \|_{F}}+\frac{\| Q_I(\dico)B_I \|^4_F}{\left(\|\dico_I^\dagger\|_{2,2}^{-1}  - 2\|B_I \|_{F}\right)^2 } \notag \\
&\leq \|\dico_I\|^2_F - \| Q_I(\dico)B_I \|^2_F + \frac{ 2 \| Q_I(\dico)B_I \|^2_F \|B_I \|_F}{\sqrt{1-\delta_S} - \|B_I \|_{F}}+\frac{\| Q_I(\dico)B_I \|^4_F}{\left( \sqrt{1-\delta_S}  - 2\|B_I \|_{F}\right)^2 }.
\end{align}
\end{proof}

\begin{lemma} \label{lem:trick}
For $a,b,\xi > 0$,
\begin{align}
\xi \leq \frac{4b}{1+ \sqrt{1+16\log(\frac{a}{b})}} \qquad \mbox{implies that} \qquad a \exp \left(\frac{-b^2}{\xi^2} \right)< \xi. 
\end{align}
\end{lemma}
\begin{proof}
We have
\begin{align*}
a \exp \left(\frac{-b^2}{\xi^2} \right)< \xi
\quad \Leftrightarrow \quad \frac{a}{b} \exp \left(-\frac{b^2}{\xi^2}\right) <  \left(\frac{b}{\xi}\right)^{-1} 
\quad \Leftrightarrow \quad \log\left(\frac{a}{b}\right) - \frac{b^2}{\xi^2} < - \log\left(\frac{b}{\xi}\right).
\end{align*}
Since $\log x < x/2$ for $x\geq 0$ the last inequality is implied by
\begin{align*}
\frac{b^2}{\xi^2}-\frac{b}{2\xi} \geq \log\left(\frac{a}{b}\right),
\end{align*}
which is satisfied as soon as 
\begin{align*}
\frac{b}{\xi} \geq \frac{1}{4} \left(1 + \sqrt{1+ 16 \log\left(\frac{a}{b}\right)} \right)  \qquad \Leftrightarrow \qquad \xi \leq \frac{4b}{1+ \sqrt{1+16\log(\frac{a}{b})}}.
\end{align*}
\end{proof}

\bibliography{/Users/karinschnass/Desktop/latexnotes/karinbibtex}

\begin{thebibliography}{10}

\bibitem{aganjaneta13}
A.~Agarwal, A.~Anandkumar, P.~Jain, P.~Netrapalli, and R.~Tandon.
\newblock Learning sparsely used overcomplete dictionaries via alternating
  minimization.
\newblock {\em arXiv:1310.7991}, 2013.

\bibitem{aganne13}
A.~Agarwal, A.~Anandkumar, and P.~Netrapalli.
\newblock Exact recovery of sparsely used overcomplete dictionaries.
\newblock {\em arXiv:1309.1952}, 2013.

\bibitem{ahelbr06}
M.~Aharon, M.~Elad, and A.M. Bruckstein.
\newblock {K}-{S}{V}{D}: An algorithm for designing overcomplete dictionaries
  for sparse representation.
\newblock {\em IEEE Transactions on Signal Processing.}, 54(11):4311--4322,
  November 2006.

\bibitem{ahelbr06b}
M.~Aharon, M.~Elad, and A.M. Bruckstein.
\newblock On the uniqueness of overcomplete dictionaries, and a practical way
  to retrieve them.
\newblock {\em Journal of Linear Algebra and Applications}, 416:48--67, July
  2006.

\bibitem{argemo13}
S.~Arora, R.~Ge, and A.~Moitra.
\newblock New algorithms for learning incoherent and overcomplete dictionaries.
\newblock {\em arXiv:1308.6273}, 2013.

\bibitem{blda08}
T.~Blumensath and M.E. Davies.
\newblock Iterative thresholding for sparse approximations.
\newblock {\em Journal of Fourier Analysis and Applications}, 14(5-6):629--654,
  2008.

\bibitem{cadedoyi06}
{E}. {C}and{\`e}s, L.~Demanet, D.L. Donoho, and L.~Ying.
\newblock Fast discrete curvelet transforms.
\newblock {\em Multiscale Modeling \& Simulation}, 5(3):861--899, 2006.

\bibitem{carota06}
{E}. {C}and{\`e}s, {J}. {R}omberg, and {T}. {T}ao.
\newblock {R}obust uncertainty principles: exact signal reconstruction from
  highly incomplete frequency information.
\newblock {\em {IEEE} {T}ransactions on {I}nformation {T}heory},
  52(2):489--509, 2006.

\bibitem{chdosa98}
S.S. Chen, D.L. Donoho, and M.A. Saunders.
\newblock Atomic decomposition by basis pursuit.
\newblock {\em SIAM Journal on Scientific Computing}, 1998.

\bibitem{ch03}
O.~Christensen.
\newblock {\em An Introduction to Frames and Riesz Bases}.
\newblock {B}irkh{\"a}user, 2003.

\bibitem{da92}
I.~Daubechies.
\newblock {\em Ten Lectures on Wavelets}.
\newblock CBMS-NSF Lecture Notes. SIAM, 1992.

\bibitem{dadefogu10}
I.~Daubechies, {R}.{A}. {D}e{V}ore, {M}. {F}ornasier, and S.~G\"{u}nt\"{u}rk.
\newblock Iteratively reweighted least squares minimization for sparse
  recovery.
\newblock {\em Communications on Pure and Applied Mathematics}, 63(1):1--38,
  January 2010.

\bibitem{damaav97}
G.~Davis, S.~Mallat, and M.~Avellaneda.
\newblock Adaptive greedy approximations.
\newblock {\em Constructive Approximation}, 13:57--98, 1997.
\newblock Springer-Verlag New York Inc.

\bibitem{do06cs}
D.L. Donoho.
\newblock {C}ompressed sensing.
\newblock {\em {IEEE} {T}ransactions on {I}nformation {T}heory},
  52(4):1289--1306, 2006.

\bibitem{doelte06}
D.L. Donoho, M.~Elad, and V.N. Temlyakov.
\newblock Stable recovery of sparse overcomplete representations in the
  presence of noise.
\newblock {\em {IEEE} {T}ransactions on {I}nformation {T}heory}, 52(1):6--18,
  January 2006.

\bibitem{olsfield96}
D.J. Field and B.A. Olshausen.
\newblock Emergence of simple-cell receptive field properties by learning a
  sparse code for natural images.
\newblock {\em Nature}, 381:607--609, 1996.

\bibitem{gewawrXX}
Q.~Geng, H.~Wang, and J.~Wright.
\newblock On the local correctness of $\ell^1$-minimization for dictionary
  learning.
\newblock {\em arXiv:1101.5672}, 2011.

\bibitem{gethci05}
P.~Georgiev, F.J. Theis, and A.~Cichocki.
\newblock Sparse component analysis and blind source separation of
  underdetermined mixtures.
\newblock {\em {IEEE} {T}ransactions on Neural Networks}, 16(4):992--996, 2005.

\bibitem{grjebaklse13}
R.~Gribonval, R.~Jenatton, F.~Bach, M.~Kleinsteuber, and M.~Seibert.
\newblock Sample complexity of dictionary learning and other matrix
  factorizations.
\newblock {\em arXiv:1312.3790}, 2013.

\bibitem{grrascva08}
R.~Gribonval, H.~Rauhut, K.~Schnass, and P.~Vandergheynst.
\newblock {A}toms of all channels, unite! {A}verage case analysis of
  multi-channel sparse recovery using greedy algorithms.
\newblock {\em Journal of Fourier Analysis and Applications}, 14(5):655--687,
  2008.

\bibitem{grsc10}
R.~Gribonval and K.~Schnass.
\newblock Dictionary identifiability - sparse matrix-factorisation via
  $l_1$-minimisation.
\newblock {\em {IEEE} {T}ransactions on {I}nformation {T}heory},
  56(7):3523--3539, July 2010.

\bibitem{bagrje13}
R.~Jenatton, F.~Bach, and R.~Gribonval.
\newblock Local stability and robustness of sparse dictionary learning in the
  presence of noise.
\newblock {\em preprint}, 2012.

\bibitem{kreutz03}
K.~Kreutz-Delgado, J.F. Murray, B.D. Rao, K.~Engan, T.~Lee, and T.J. Sejnowski.
\newblock Dictionary learning algorithms for sparse representation.
\newblock {\em Neural Computations}, 15(2):349--396, 2003.

\bibitem{krra00}
K.~Kreutz-Delgado and B.D. Rao.
\newblock {FOCUSS}-based dictionary learning algorithms.
\newblock In {\em SPIE 4119}, 2000.

\bibitem{leta91}
M.~Ledoux and M.~Talagrand.
\newblock {\em {P}robability in {B}anach spaces. {I}soperimetry and processes.}
\newblock {S}pringer-{V}erlag, {B}erlin, {H}eidelberg, {N}ew{Y}ork, 1991.

\bibitem{mabaposa10}
J.~Mairal, F.~Bach, J.~Ponce, and G.~Sapiro.
\newblock Online learning for matrix factorization and sparse coding.
\newblock {\em Journal of Machine Learning Research}, 11:19--60, 2010.

\bibitem{mapo10}
A.~Maurer and M.~Pontil.
\newblock {K}-dimensional coding schemes in {H}ilbert spaces.
\newblock {\em {IEEE} {T}ransactions on {I}nformation {T}heory},
  56(11):5839--5846, 2010.

\bibitem{megr12}
N.A. Mehta and A.G. Gray.
\newblock On the sample complexity of predictive sparse coding.
\newblock {\em arXiv:1202.4050}, 2012.

\bibitem{cosamp}
D.~Needell and {J}.{A}. {T}ropp.
\newblock {C}o{S}a{MP}: Iterative signal recovery from incomplete and
  inaccurate samples.
\newblock {\em Applied Computational Harmonic Analysis}, 26(3):301--321, 2009.

\bibitem{pl07}
M.D. Plumbley.
\newblock Dictionary learning for $\ell_1$-exact sparse coding.
\newblock In M.E. Davies, C.J. James, and S.A. Abdallah, editors, {\em
  International Conference on Independent Component Analysis and Signal
  Separation}, volume 4666, pages 406--413. Springer, 2007.

\bibitem{cs}
DSP Rice~University.
\newblock Compressive sensing resources.
\newblock {\em http://www.compressedsensing.com/}.

\bibitem{rubrel10}
R.~Rubinstein, A.~Bruckstein, and M.~Elad.
\newblock Dictionaries for sparse representation modeling.
\newblock {\em Proceedings of the IEEE}, 98(6):1045--1057, 2010.

\bibitem{sc13sampta}
K.~Schnass.
\newblock Dictionary identification results for {K-SVD} with sparsity parameter
  1.
\newblock In {\em SampTA13}, 2013.

\bibitem{sc13arxiv}
K.~Schnass.
\newblock On the identifiability of overcomplete dictionaries via the
  minimisation principle underlying {K-SVD}.
\newblock {\em arXiv:1301.3375}, 2013.

\bibitem{scva07}
K.~Schnass and P.~Vandergheynst.
\newblock Average performance analysis for thresholding.
\newblock {\em IEEE Signal Processing Letters}, 14(11):828--831, 2007.

\bibitem{sken10}
K.~Skretting and K.~Engan.
\newblock Recursive least squares dictionary learning algorithm.
\newblock {\em {IEEE} {T}ransactions on {S}ignal {P}rocessing},
  58(4):2121--2130, April 2010.

\bibitem{spwawr12}
D.~Spielman, H.~Wang, and J.~Wright.
\newblock Exact recovery of sparsely-used dictionaries.
\newblock In {\em Conference on Learning Theory (arXiv:1206.5882)}, 2012.

\bibitem{Tropp:greed}
{J}.{A}. {T}ropp.
\newblock Greed is good: Algorithmic results for sparse approximation.
\newblock {\em IEEE Transactions on Information Theory}, 50(10):2231--2242,
  October 2004.

\bibitem{tr08}
{J}.{A}. {T}ropp.
\newblock On the conditioning of random subdictionaries.
\newblock {\em Applied Computational Harmonic Analysis}, 25(1-24), 2008.

\bibitem{vamabr11}
D.~Vainsencher, S.~Mannor, and A.M. Bruckstein.
\newblock The sample complexity of dictionary learning.
\newblock {\em Journal of Machine Learning Research}, 12(3259-3281), 2011.

\bibitem{ve10}
R.~Vershynin.
\newblock Introduction to the non-asymptotic analysis of random matrices.
\newblock In Y.~Eldar and G.~Kutyniok, editors, {\em Compressed Sensing, Theory
  and Applications}, chapter~5. {C}ambridge {U}niversity {P}ress, 2012.

\bibitem{yablda09}
M.~Yaghoobi, T.~Blumensath, and M.E. Davies.
\newblock Dictionary learning for sparse approximations with the majorization
  method.
\newblock {\em {IEEE} {T}ransactions on {S}ignal {P}rocessing},
  57(6):2178--2191, June 2009.

\bibitem{zipe01}
M.~Zibulevsky and B.A. Pearlmutter.
\newblock Blind source separation by sparse decomposition in a signal
  dictionary.
\newblock {\em Neural Computations}, 13(4):863--882, 2001.

\end{thebibliography}
\bibliographystyle{plain}
\end{document}